\documentclass[aps,pra,floatfix,nofootinbib]{revtex4-2}
\usepackage{amsmath}
\usepackage{graphicx}
\usepackage{amssymb}
\usepackage{amsmath}
\usepackage{xcolor}

\usepackage{graphicx,graphics,booktabs,mathtools,color,setspace}

\usepackage{accents}
\newlength{\dhatheight}

\usepackage{url}

\newcommand{\half}{\ensuremath{\textstyle\frac{1}{2}}}
\newcommand{\quarter}{\ensuremath{\textstyle\frac{1}{4}}}
\newcommand{\rmd}{\mathrm{d}}
\newcommand{\rme}{\mathrm{e}}
\newcommand{\rmi}{\mathrm{i}}

\begin{document}

\title{Multi-spin control from one-spin pulses}
\author{Suzanne Lim, Bowen Guo, Abi Turner, Charles Buchanan}
\affiliation{Chemistry Research Laboratory, University of Oxford, Mansfield Road, Oxford OX1 3TA, UK}
\affiliation{Kavli Institute for Nanoscience Discovery, University of Oxford, Oxford, UK}
\author{Jonathan A. Jones}
\affiliation{Clarendon Laboratory, University of Oxford, Parks Road, Oxford OX1 3PU, UK}
\author{Andrew Baldwin}
\email[Correspondence should be addressed to ]{andrew.baldwin@chem.ox.ac.uk}
\affiliation{Chemistry Research Laboratory, University of Oxford, Mansfield Road, Oxford OX1 3TA, UK}
\affiliation{Kavli Institute for Nanoscience Discovery, University of Oxford, Oxford, UK}

\date{\today}

\begin{abstract} Controlling ensembles of weakly coupled spins typically requires computationally expensive multi-spin optimisations. We present a compact framework that enables control of weakly coupled spin systems (of any spin), but using RF pulses optimised for a single spin-\half. We do this by explicitly creating a GRAPE pulse with fixed `active' evolution times using single spin-$\frac{1}{2}$ methods, and pulsing on one spin at a time. By enforcing this form uniformly across offsets (`band-schematic' pulses), chemical shift and scalar coupling evolution of the entire system can be precisely controlled. We demonstrate the approach by constructing band-schematic pulses and a continuously irradiated joint INEPT (JINEPT) that achieves band-selective transfer $I_z\rightarrow 2I_zS_z$. The framework is implemented in the software Seedless, which both rapidly generates such pulses and analyses the schematic form of arbitrary pulses, enabling robust multi-spin control, without multi-spin optimisation. \end{abstract}

\maketitle

\doublespacing

\section{Introduction} Nuclear magnetic resonance (NMR) experiments rely on sequences of radiofrequency (RF) pulses and delays to control spin dynamics and encode molecular information. In most applications, RF pulses must act uniformly across a range of resonance offsets and tolerate spatial variations in the $B_1$ field. Rectangular pulses are simple but limited in bandwidth and robustness, motivating the widespread use of shaped pulses designed by optimal control methods such as GRadient Ascent Pulse Engineering (GRAPE)~\cite{glaser1998,skinner2003,khaneja2005,gershenzon2008,skinner2012,Braun2014}.

Although GRAPE can be applied to arbitrary spin systems~\cite{ehni2022}, the computational cost grows rapidly with system size, making full multi-spin optimisation impractical for routine or adaptive use. By contrast, pulses optimised for a single spin-\half\ can be computed both extremely rapidly and reliably, and are therefore attractive for `on-the-fly' production. By  accounting for the $B_1$ inhomogeneities and bandwidth requirements for a specific combination of sample and hardware, the sensitivity of commonly used experiments can be increased substantially~\cite{seedless}. 

It is well known that rectangular pulses when used for a 90$^\circ$ rotation near-resonance lead to a $\frac{2}{\pi}T$ evolution that occurs before and after the action of the pulse~\cite{hoult1979,bain2009}. These can be subtracted from adjacent delays in applications where chemical shift or scalar coupling evolution times need to be precisely controlled. This is critical in, for example, triple resonance experiments~\cite{kay1990}. Generalising this, Lescop \textit{et al}.\ noted that the action of many shaped pulses can be described by a `schematic' form, consisting of free evolution periods surrounding a central rotation~\cite{schanda2006, lescop2007, lescop2010}, whose form is maintained over a range of chemical shifts. These delays can be subtracted, a finding critical for the BEST-TROSY experiments~\cite{schanda2006, lescop2007, lescop2010}. GRAPE pulses have been created using single spin methods by exploiting this schematic form for control of scalar coupling evolution in multi-spin ensembles~\cite{gershenzon2008,Braun2014}, and have been used to create co-operative pulses, where the evolution delays of a series of pulses are controlled to optimise transfers~\cite{Braun2014}. Here we focus on individual pulses, extend the formalism underlying this concept and develop methods to exploit it systematically, allowing tailored evolution controlled pulses to be produced rapidly by spectroscopists. 

We show that provided only one spin is excited at a time, then the evolution of weakly-coupled spin systems with nuclei of any spin, of arbitrary size and connectivity can be precisely controlled using pulses optimised for an isolated spin-$\frac{1}{2}$, with schematic parameters that are uniform across offsets (`band-schematic'). We present a general method to extract the schematic description of arbitrary pulses and use this to analyse common classes of pulses to determine their optimal use. We demonstrate this by developing the joint INEPT (JINEPT) transfer element, where we transfer polarisation between spins under constant irradiation, and show that varying delays in an INEPT contructed with rectangular pulses is equivalent to varying the evolution parameters in a band-schematic pulse. This framework enables precise control of chemical shift and scalar coupling evolution in multi-spin sequences while retaining the speed and flexibility of single-spin optimal control. These methods have been implemented in Seedless, enabling pulses for common applications to be generated in under a second on a modern laptop, allowing them to be tuned for any hardware/sample requirement to enhance sensitivity in NMR experiments.

\section{Results}

\subsection{Controlling ensembles of spins, using single spin pulses}
We want to control the evolution of all spins within a weakly coupled ensemble. To do this (for derivation, Appendix \ref{isomorph}), we ensure that we only pulse on one spin at a time, and create a pulse acting on a single spin that follows a schematic form~\cite{gershenzon2008,lescop2010},
\begin{equation} V_\mathrm{S} = Z(b\Omega T)\,U\,Z(a\Omega T), 
\label{schem} 
\end{equation}
where $U$ is an instantaneous rotation about an arbitrary axis/angle, $aT$ and $bT$ are pre- and post-evolution times, expressed as fractions of the pulse duration $T$, and $Z(\theta)$ describes a rotation of specified angle about the $Z$ axis. When used on a weakly coupled system of spin-\half, both chemical shift of the pulse spin, and all scalar couplings to that spin will evolve for the pre- and post- evolution periods provided that $U$, $a$, and $b$ are the same at both $\Omega\pm\frac{1}{2}J$. More generally, the pulse is \emph{band-schematic} when $U$, $a$ and $b$ are constant over a range of offsets that is wide enough to control the evolution frequencies of all spins of interest, including couplings. The chemical shift evolution of all other spins, and scalar coupling evolution of all couplings to all nuclei other than the one being excited will evolve for the full duration $T$. Within a pulse sequence, the total evolution periods can be  controlled by adjusting the duration of flanking delays, allowing precise control of multi-spin evolution.

This picture is not restricted to spin-$\frac{1}{2}$. The computation of a pulse for a nucleus of any spin will be isomorphic to the spin-$\frac{1}{2}$ case provided all anisotropic interactions such as quadrupolar couplings are motionally averaged to zero, so that the Hamiltonian has terms with linear operators and weak scalar couplings only (Appendix \ref{isomorph}). Provided there is no strong coupling and that a pulse sequence applies RF radiation to just one spin at a time, the exact duration for which scalar coupling and chemical shift are active over a range of chemical shifts, for arbitrarily sized systems for any value of spin can be exactly controlled using evolution-controlled pulses designed for a single spin-\half. 

Some, but not all commonly used shaped pulses have a schematic form close to this (section \ref{sec:example}). Band-schematic single-spin pulses with $B_1$ field robustness can be generated by specifying the schematic propagator $V_\mathrm{S}$ as the target as a `universal' rotation restraint, within a GRAPE framework to \textit{exactly} match a required application. Five parameters are required to define $V_\mathrm{S}$: the axis ($u_\phi,u_\theta$) and angle ($u_\psi$) of $U$, and the fractional evolution delays $a$ and $b$. 

Three methods have been included in Seedless to create these (Appendix \ref{app:SeedScript}). A target operator such as \verb|a90xb| can be supplied as a target, and the evolution delays can be set using \verb|evAlpha=a| and \verb|evBeta=b| in the header. The values of $a$ and $b$ typically need to be in the range 0--1, and can be adjusted to to obtain an acceptably low optimised infidelity score. Alternatively, an explicit unitary operator can be constructed and used as a target such as \verb|bZ;0.25B;aZ|, where a semi-colon denotes separate rotations (it is rarely necessary to use these explicit forms as the standard targets are usually sufficient, but a full description of the general syntax is in the Seedless manual). It is also possible to create state-to-state pulses that are semi-evolution controlled, by specifying, \textit{e.g.}, \verb|Iz -bOIy| as the initial and final states (where $b$ can be zero). This leads to one of the two evolution delays (in this case, the second) being controlled. 

There are singularities in $V_\mathrm{S}$ for some parameters, including 180$^\circ$ pulses. Evolution-controlled pulses that perform this action can still be constructed, but modifications are needed to the approach (section \ref{sec:180x}). Overall, these approaches preserve the computational efficiency of single-spin optimal control while enabling explicit management of evolution effects in multi-spin experiments.

\subsection{Determining the schematic form of arbitrary pulses}
To use conventional shaped pulses reliably in multi-spin sequences, we need to determine their schematic parameters and how these vary with offset. If we know the axis of the central rotation $U$, we could perform a Euler decomposition of the pulse propagator $V$ (e.g. about the ZXZ axes if $U$ is aligned along $x$) to obtain 3 rotation angles~\cite{Braun2014}. In a general case, this axis is not known, and so there is insufficient information in the propagator to obtain the 5 schematic parameters. Effective evolution delays $a$ and $b$ have been computed for some common pulse shapes using two-spin simulations of INEPT transfers~\cite{lescop2010} which also require knowledge of the axis of rotation. We develop here a single-spin method to provide the unique mapping between the action of a pulse, and its schematic form.

We note that the schematic form for infinitesimal evolution periods already resembles the Trotter--Suzuki form (Appendix~\ref{app:opdev}). To reliably extract the schematic pulse parameters outside of this limit, we take a lead from Feynman and Waugh~\cite{Feyn1951,wilcox1967,waugh1968}, and note that an arbitrary pulse with propagator $V(\Omega)$ has a one-sided exact decomposition  $V = U P$ (Appendix~\ref{app:opdev}), where we define a rotation $P=\rme^{-\rmi T p \Omega}$ that captures the net linear evolution experienced during the pulse. We can calculate the dimensionless Hermitian evolution operator $p$ via 
\begin{equation} 
p= \frac{\rmi}{T}V^\dagger \frac{\rmd V}{\rmd \Omega}.
\label{eq:pdef}
\end{equation} 
For a pulse to be band-schematic, we require $p$ to be independent of $\Omega$ over a frequency range, from which it follows that $U$ is also independent of $\Omega$ (Appendix~\ref{app:opdev}). In this case, all evolution information is described by $p$. For an arbitrary pulse for a given $\Omega$, both $V$ and $p$ can be calculated (Appendix~\ref{app:calcP}). From these, the five schematic parameters $(a,b,u_\psi,u_\theta,u_\phi)$ can be recovered analytically (Appendix~\ref{app:calcSchem}). 

\begin{figure}
\includegraphics[trim={1cm 21.5cm 0 0},width=1.05\linewidth]{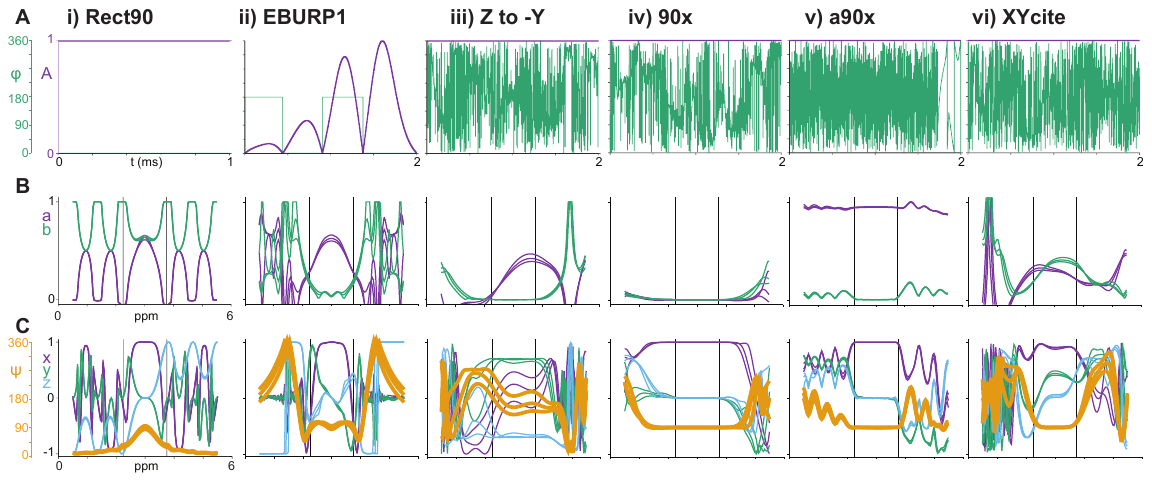}
\caption{The schematic form of selected excitation/de-excitation pulses (Table \ref{tab1}). \textbf{A} The normalised amplitude (A) and phase ($\phi$) of a i) rectangular $90^\circ$ (1\,ms, $\omega$ 0.25\,KHz), ii) an EBURP1~\cite{burp1991} (2\,ms, $\omega$ 1.86\,KHz, giving a 3\,ppm bandwidth) that has approximately consistent schematic parameters over its range of activity, and four pulses generated using Seedless (2\,ms, $\omega$ 5\,KHz, optimised between $-1.5$ and 1.5\,ppm, indicated by vertical bars, with a $B_1$ field distribution of 0.95, 1, 1.03 with constant amplitude~\cite{seedless}, Appendix \ref{app:SeedScript}),  iii) a $Z\rightarrow-Y$ state to state showing `semi' evolution control ($a$ is uncontrolled and $b=0$), iv) a 90x (where $a=b=0$), v) an evolution controlled a90xb (where $a=0.95$, $b=0$) and vi) an XYcite where $a$ and $b$ are uncontrolled. The infidelities of the optimised Seedless pulses were below 10$^{-4}$ indicating that the pulses well perform their intended function. The schematic pulse parameters of each were computed versus offset (computed for a Larmor frequency of 600\,MHz). \textbf{B} The pre ($a$, purple) and post ($b$, green) evolution delays of each are indicated. \textbf{C} The angle ($\psi$, orange) and unit vector ($x,y,z$) describing the central rotation versus $\Omega$. Only the 90x and a90x are band-schematic in the range optimised with $a$ and $b$ staying constant over the region. In all plots, the performance of the pulses are shown at $B_1$ valuse of 0.95, 1, 1.03 of the central value. In the general case, a new pulse can be rapidly constructed using Seedless~\cite{seedless} for any of these tasks, as described in the text (Appendix \ref{app:SeedScript}).}
\label{fig:90}
\end{figure}

It is instructive to take an example. For a rectangular $90^\circ_x$ pulse near resonance (Figure \ref{fig:90}i), $V\approx\rme^{-\rmi \frac{\pi}{2} I_x}$. Evaluating $p$ (Appendix~\ref{app:calcP}) and projecting both $V$ and $p$ onto the Cartesian basis $I=(I_x,I_y,I_z,E)$ yields $\Vec{V}=(-\rmi\sqrt{2} ,0 ,0,\frac{1}{\sqrt{2} })$ and $\Vec{p}=( 0 , \frac{2}{\pi},\frac{2}{\pi},0)$. Applying Equation \ref{eq:decomp} yields $a=b=\frac{2}{\pi}$ with $90_x$ as the central rotation ($u_\phi=0$, $u_\theta=u_\psi=\frac{\pi}{2}$), as expected. The need to account for the pre and post $\frac{2}{\pi}t$ evolution periods associated with rectangular pulses is well known~\cite{hoult1979,bain2009,kay1990}.

After either creating or reading in a pulse in Seedless, specifying \verb|EVOLVE| in the input file will generate plots showing how these parameters vary with offset (Figures \ref{fig:90} and \ref{fig:180}, Appendix \ref{app:SeedScript}). This procedure enables quantitative analysis of arbitrary shaped pulses using single-spin calculations, providing direct insight into their suitability for use in multi-spin pulse sequences.

\begin{table}
  \begin{tabular}{lc|cc|ccc|cc|cc}
\textbf{Pulse} & \textbf{GRAPE} &\textbf{Axis} &  \textbf{$\psi$} & \textbf{$a$} & \textbf{$b$} & \textbf{$c$} & \textbf{Half: Axis} & \textbf{$\psi$} &  \textbf{$a$ control?} & \textbf{$b/c$ control?}\\
    \hline
    rect.90$^\circ$    & n & (1,0,0) & 90$^\circ$ &  0.64 &  0.64 & &  (1,0,0) & 45$^\circ$ &   No & No \\
   EBURP1\cite{burp1991}              & n & (1,0,0) & 90$^\circ$ &   0.63  & 0.03 &  & (1,0,0) & 28$^\circ$ &   Partial & Partial \\
    $Z\rightarrow -Y$  & y & (0.25,0.68,-0.68) & 152$^\circ$ &  0.33 & 0.00  & & (0.19,0.75,0.63) &  232$^\circ$ &   No & Yes \\
    \verb|90x|         & y & (1,0,0) & 90$^\circ$ &  0.00 & 0.00   & & (-0.88,-0.48,-0.01) & 167$^\circ$ &   Yes & Yes \\
    \verb|a90x|        & y & (1,0,0) & 90$^\circ$ &  0.95 & 0.00  & & (0.40,-0.22,-0.89) & 6$^\circ$ &   Yes & Yes \\
    \verb|XYcite|      & y & (0.9,0.4,0.04) & 90$^\circ$ & 0.3 & 0.4 & & (0.69,-0.21,-0.69) & 93$^\circ$ & No & No \\
    \verb|PC9|\cite{kupce1994}        & n & (1,0,0) & 90$^\circ$ &  0.06 & 0.95  & &  (1,0,0)    & 83.2$^\circ$    &    Yes & Yes \\
    \verb|BESTPC9|\cite{schanda2006}     & n & (1,0,0) & 90$^\circ$ & 0.51 & 0.51  & &   (1,0,0)  & 45$^\circ$      &  Yes &  Yes \\
    \verb|Q5|\cite{emsley1992}          & n & (1,0,0) & 90$^\circ$ & 0.01 & 0.08  & &   (-1,0,0) & 194$^\circ$     &  No  & No  \\
    \hline               
    rect.180$^\circ$   & n & (1,0,0) & 180$^\circ$ & 0.32   & 0.64 & 0.32 & (1,0,0) & 90$^\circ$ &   No & No \\
   REBURP\cite{burp1991}     & n& (1,0,0) & 180$^\circ$ & 0.47   & 0 & 0.47   & (1,0,0) & 90$^\circ$ &   Yes & Yes \\
    \textbf{$Z \rightarrow -Z$}  & y & (0.07,-1.00,0.00) & 180$^\circ$ & 0.26   & 0.07 & 0.26 & (0.61,0.47,0.64) & 226$^\circ$ &   No & No \\
   \verb|180x|         & y & (1,0,0) & 180$^\circ$ & 0.13  &  0.12& 0.14  & (0.34,-0.9,-0.27) & 90$^\circ$ &   No & No \\
    \verb|a180xa|      & y & (1,0,0) & 180$^\circ$ &  0.47 & 0  & 0.47 & (1,0,0) & 90$^\circ$ &   Yes & Yes \\
    \verb|Q3|\cite{emsley1992}          & n & (1,0,0) & 180$^\circ$ & 0.31 & 0.07 & 0.19 & (-1,0,0) & 150$^\circ$ & No & No \\
    CHIRP\cite{chirp1989} & n & (0.13,0.99,0) & 180$^\circ$ & 0.35 & 0.02 & 0.35 & (0.7,-0.1,-0.7) & 183$^\circ$ & No & No \\
  \end{tabular}
\caption{The on-resonance schematic characteristics of a series of excitation/de-excitation pulses (top) and 180$^\circ$ rotations (bottom), evaluated near resonance. Selected pulses from this table are shown with their variation versus $\Omega$ and $B_1$ robustness (Figures \ref{fig:90} and \ref{fig:180}). The non-GRAPE pulses were calculated using Seedless following instructions from their original publications (full details in Appendix \ref{app:SeedScript}). Phase-only GRAPE pulses created by Seedless for this work are indicated, and their performance and optimisation criteria are shown (Figures \ref{fig:90} and \ref{fig:180}, Appendix \ref{app:SeedScript}). Both (90x, a90x, a180xa), one ($Z\rightarrow -Y$), or neither (180x, XYcite, $Z\rightarrow -Z$) of the pre/post evolution delays are band-schematic in these optimisations, as described in the text (indicated by the final 2 columns, where `partial' indicates variations $>10\%$ with offset). The axis and angle of the central rotation at full time is shown, together with the action of the front half of the pulse ($U_1$), as described in the text. For the excitation/de-excitation pulses, we need only a pre ($a$) and post ($b$) fractional evolution delay to understand their action. The evolution periods of the 180$^\circ$ pulses were analysed by breaking the pulse two at the half way point and computing, with the pre ($a$), middle ($b$) and post ($c$) fractional evolution delays shown, as described in the text. All pulses indicated perform the required overall transform on the density matrix. Only a subset of the pulses do so while keeping evolution delays constant over a range of $\Omega$. In the general case, a new pulse can be rapidly constructed using Seedless~\cite{seedless} for any of these tasks, as described in the text (Appendix \ref{app:SeedScript}).}
\label{tab1}
\end{table}

\noindent

\subsection{Schematic description of commonly used pulse types}
\label{sec:example}
We discuss how commonly used pulse classes can be created, and how they relate to the schematic picture (Figures \ref{fig:90},\ref{fig:180}, Table \ref{tab1}). All pulses discussed here, both GRAPE and non-GRAPE, can be computed using Seedless (Appendix \ref{app:SeedScript}). In what follows, we make use of the fact that we can make a new pulse from an original by reversing its order and phase~\cite{levitt1982,PulseSymNgoMorris1987,luy2005,skinner2012} (Appendix \ref{app:timePhaseRev}). If the original propagator has a schematic form, then the propagator of the new pulse will be
\begin{equation}
V^\prime_S = Z(a\Omega T) U^\prime Z(b\Omega T)
\end{equation}
where the new central unitary is related to the first by $U^\prime = Y(\pi) U^\dagger Y^\dagger(\pi)$. If the rotation is in the $XZ$ plane, then $U=U^\prime$ and creating the order/phase reversed pulse simply swaps the pre- and post- evolution delays (other symmetry partners can also be created, Table \ref{tab:sym}). This allows pulses with a schematic form to be used in pairs, flanking a central 180$^\circ$ pulse, where evolution created by the first can be exactly undone by the second. The additional evolution periods must be accounted for in pulse sequences when used outside purely longitudinal magnetisation.

Finally, owing to singularities in $V_\mathrm{S}$, not all evolution-controlled pulses can be generated directly by specifying $V_\mathrm{S}$, and alternative approaches are described.

\subsubsection{Evolution controlled 90$^\circ_x$ pulses}
For unitary $90^\circ$ pulses with evolution control, \textit{e.g}.\ \verb|a90xb| (Figure \ref{fig:90}v), both parameters $a$ and $b$ are linearly independent in $V_\mathrm{S}$, and can be controlled. In a general case it will be advantageous to use pulses created using $V_\mathrm{S}$ using Seedless to ensure evolution control.

We consider the degree to which previously published pulses are band-schematic. $90_x$ pulses have been produced~\cite{seedless}, which effectively have $a=b=0$ (Figure \ref{fig:90}iv). Rectangular $90^\circ_x$ pulses are not band-schematic (Figure \ref{fig:90}i), but close to resonance are described by $a=b=\frac{2}{\pi}$ and $U=X(\frac{\pi}{2})$. The widely used EBURP1~\cite{burp1991} is approximately, but not exactly band-schematic (Figure \ref{fig:90}ii) as over its range of action, the pre- and post- evolution periods vary. Near resonance it has the schematic form~\cite{lescop2010}
\begin{equation}
V_{\mathrm{EBURP1}}= Z(0.03\Omega T) X(\frac{\pi}{2}) Z(0.62 \Omega T)   
\end{equation}

The PC9 pulse ($V_{\mathrm{PC9}}$)~\cite{kupce1994} optimised for a $90^\circ$ rotation is also band-schematic, with a pre- and post- evolution delay. The version adapted for use in the BEST-TROSY experiments~\cite{schanda2006,lescop2007,lescop2010} concatenates this with a time reversed symmetry partner ($\overline{V_{\mathrm{PC9}}}$, Appendix~\ref{app:timePhaseRev}) leading to 
\begin{equation}
   V_{\mathrm{BESTPC9}}= V_{\mathrm{PC9}}\overline{V_{\mathrm{PC9}}}=Z(0.51\Omega T)X(\frac{\pi}{2})Z(0.51\Omega T),
\end{equation}
where the maximum amplitude occurs at the centre of the pulse. This is both band-schematic and has excellent $B_1$ robustness as elegantly exploited in the BEST-TROSY experiments~\cite{schanda2006,lescop2007,lescop2010}. The \verb|Q5|~\cite{emsley1992} is also widely used, but is not band-schematic and when used in multi-pulse sequences, evolution period under scalar coupling will vary with $\Omega$. In the general case, Seedless can produce pulses that are band-schematic by design (Appendix \ref{app:SeedScript}).

\subsubsection{State-to-state excitation and de-excitation}
\label{sec:ss}
State-to-state excitation and de-excitation pulses control one of the two evolution delays, and can be used effectively when the side with uncontrolled evolution is applied to $Z$ magnetisation. For example, an excitation (e.g. $Z\rightarrow -Y$, Figure \ref{fig:90}iii) pulse will have the schematic form (Appendix~\ref{app:timePhaseRev})
\begin{equation}
\begin{split}
V_{Z\rightarrow -Y}= U Z(a\Omega T), \\
\end{split}
\label{eq:schemSS}
\end{equation}
where $U$ could be one of a set of rotations that cause the transform, whose axis is equidistant to the start and finish positions on the Bloch sphere, the post-evolution delay $b$ is fixed at zero, and the pre-evolution delay $a$ need not be controlled. A new pulse can be created by order/phase reversal~\cite{levitt1982,PulseSymNgoMorris1987,luy2005,skinner2012} that performs $Y\rightarrow Z$ (Appendix~\ref{app:timePhaseRev}), with schematic form
\begin{equation}
\begin{split}
V^\prime_{Z\rightarrow-Y}=V_{Y\rightarrow Z}= Z(a\Omega T) U^\prime. \\
\end{split}
\label{eq:schemSS2}
\end{equation}
The pre-evolution delay is now a post-evolution delay and the acting unitary is related by symmetry to the original. A state-to-state pulse can also be created to perform
\begin{equation}
 I_z\rightarrow  -I_y\cos(b \Omega T) + I_x\sin(b \Omega T),
 \end{equation}
which is the combination of an $I_z\rightarrow -I_y$ excitation followed by evolution for a duration $bT$. The resulting pulse has a schematic form where the post evolution delay $b$ is controlled and hence band-schematic, but the pre-evolution delay is not and setting $b=0$ recovers the earlier result. The syntax in Seedless to create this pulse as a state-to-state transform is \verb| Iz -0.2OIy| where the value 0.2 will be interpreted as the value $b$ in the above expression (Appendix \ref{app:SeedScript}). 

\subsubsection{XYcite}
XYcite pulses (Figure \ref{fig:90}vi) are state-to-state pulses designed to get magnetisation into the XY plane but without phase alignment. They have the most general schematic form 
\begin{equation}
V_{\mathrm{XYcite}} = Z(b\Omega T) U Z(a\Omega T)
\end{equation}
where neither $a$, $b$ nor $U$ are band-schematic and the central rotation is not restrained to be a $90^\circ$ rotation. They have one advantage in that they can be short, 
but when used in a multi-spin context they can introduce frequency dependent evolutions that can cause different spins to evolve under scalar coupling for variable lengths of time. The evolution induced by one can be compensated by using them with order/phase reversed pairs.

\begin{figure}
\includegraphics[trim={1cm 16.5cm 0 0},width=1.00\linewidth]{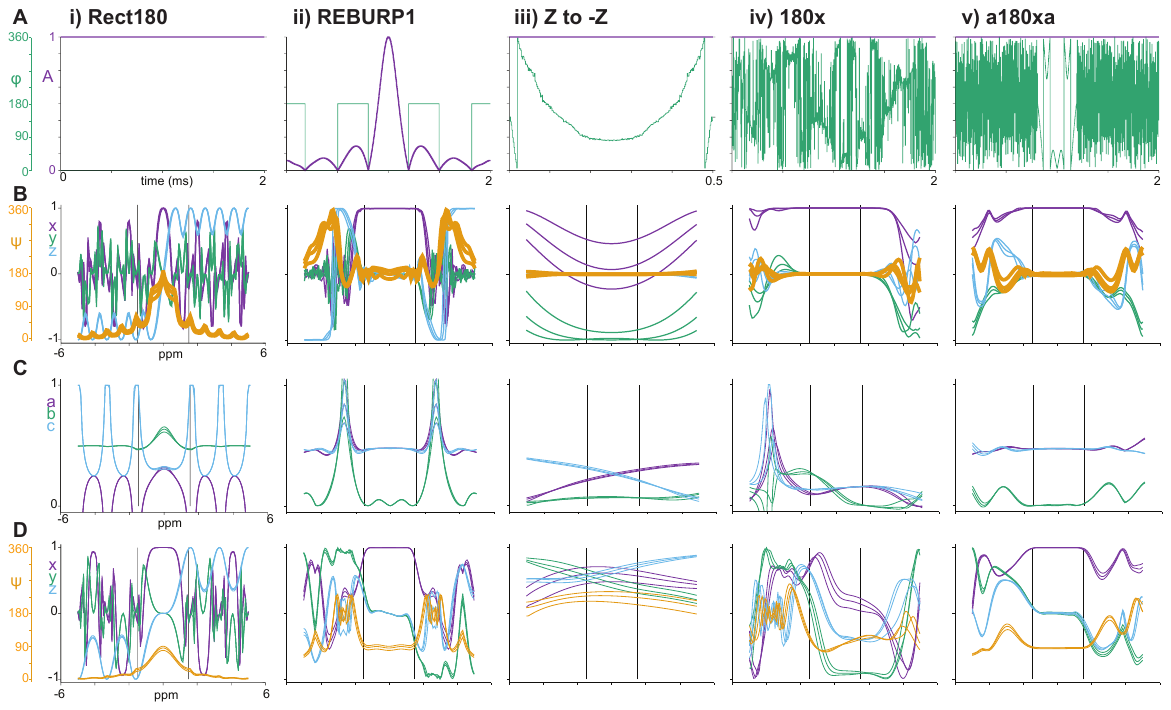}
\caption{The schematic form of selected $180^\circ$ pulses (Table \ref{tab1}). \textbf{A} The normalised amplitude (A) and phase ($\phi$) of i) a rectangular $180^\circ$ (2\,ms, $\omega$ 0.25\,KHz), ii) a REBURP1~\cite{burp1991} (2\,ms, $\omega$ 3.13\,KHz giving a 3\,ppm bandwidth), and four pulses generated using Seedless ($\omega$ 5\,KHz, optimised between $-1.5$ and 1.5\,ppm, indicated by vertical bars sampled at 0.95, 1.0 and 1.03 $B_1$ at constant amplitude~\cite{seedless}, Appendix \ref{app:SeedScript}),  iii) a $Z\rightarrow-Z$ state to state (0.5\,ms), iv) a 180x (2\,ms), and v) an evolution controlled a180xa (2\,ms). The infidelities of the Seedless pulses were below 10$^{-4}$ indicating that the pulses well perform their intended function. The schematic pulse parameters of each were computed versus offset (computed for a Larmor frequency of 600\,MHz). The pre- and post- evolution delays of the schematic description are unreliable for the case of a $180^\circ$ rotation as described in the text, and the pulses are analysed from breaking them into two halves. \textbf{B} The angle ($\psi$, orange) and axis ($x,y,z$) of the central rotation of the pulse overall. \textbf{C} The pre- ($a$, purple) middle ($b$, green) and post ($c$, blue) evolution delays derived from analysing the pulses in two halves. \textbf{D} The angle ($\psi$, orange) and axis ($x,y,z$) of the central rotation in the first half of the pulse. The overall unitary transformation reveals all pulses perform a $180^o$ rotation in their active band, but the axis varies. Only the REBURP and the a180xa pulses are band-schematic and suitable for refocusing, with pre and post $a$ and $c$ delays constant versus ppm, the mid delay $b=0$ (or close to zero for the REBURP1) and a $90_x$ for the central rotation of the first half of the pulse. The $Z\rightarrow -Z$ has uncontrolled pre- and post- evolution delays but the central evolution period $b=0$. This isn't important if the pulse is used for inversion as magnetisation will be longitudinal during the pre- and post- evolution delays. The 180x is uncontrolled for all evolution periods and so is unsuitable for refocusing (unless it is very short). In all plots, the performance of the pulses are shown at $B_1$ values of 0.95, 1, 1.03 of the central value. In the general case, a new pulse can be rapidly constructed using Seedless~\cite{seedless} for any of these tasks, as described in the text (Appendix \ref{app:SeedScript}).}
\label{fig:180}
\end{figure}

\subsubsection{Evolution controlled 180$^\circ$ pulses}
\label{sec:180x}
Refocusing pulses require more care: $V_\mathrm{S}$ for a $180^\circ$ schematic pulse in the $XY$ plane forms a singularity that does not permit independent control of $a$ and $b$ (Appendix~\ref{app:calcSchem}). We cannot create an evolution controlled 180$^o$ pulse through direct application of $V_\mathrm{S}$, and pulses created with the propagator $180_x$ will have  uncontrolled values of $a$ and $b$ (Figure \ref{fig:180}iv). 

We can create an \verb|a180xa| evolution controlled refocusing pulse by first creating an \verb|a90xb| (setting $b=0$ to generate $V_{\mathrm{a90x}}$) and then computing its order/phase reversed partner ($V_{\mathrm{a90x}}^\prime$, Appendix \ref{app:timePhaseRev}) and combining the two (Figure \ref{fig:180}iv), leading to the desired schematic form
\begin{equation}
\begin{split}
V_{\mathrm{a180xa}}=V_{\mathrm{a90x}}^\prime V_{\mathrm{a90x}} &=   Z(a \Omega T)  X(\frac{\pi}{2})   X(\frac{\pi}{2}) Z(a \Omega T)  \\
 &=    Z(a\Omega T)  X(\pi)  Z(a\Omega T).  \\
\end{split}
\end{equation}

An analysis of the commonly used REBURP pulse~\cite{burp1991} reveals that to a good approximation it's schematic form~\cite{lescop2010}is (Figure \ref{fig:180}ii)
\begin{equation}
V_{\mathrm{REBURP}}=Z(0.475 \Omega T) X(\pi) Z(0.475\Omega T).
\end{equation}
Specifying \verb|REBURP| in the Seedless input file and \verb|180x| as the target propagator will generate an $\verb|a180xa|$ pulse. The pulse will be scored twice, firstly against \verb|a90x| at the half way stage, and then against \verb|180x| overall (Figure \ref{fig:180}iv, Appendix \ref{app:SeedScript}).

Evolution controlled 180$^o$ pulses need to be analysed using a similar method. Specifying \verb|HALF| in the Seedless input file will lead to an additional schematic analysis of both the back, and front halves of the pulse (Figure \ref{fig:180}C/D). The report will show the variation of two effective propagators $U_1$ and $U_2$ and the pre-, middle- and post- evolution periods ($a,b,c$, where $b$ is the merged evolution period of  the end of the front half and the beginning of the back half) with $\Omega$, according to
\begin{equation}
V=V_{s2}.V_{s1}=Z(c\Omega T)\, U_2 \, Z(b\Omega T) \, U_1 \, Z(a\Omega T)
\end{equation}
An idealised evolution controlled \verb|a180xa| pulse has $a=c$, $b=0$ and $U_1=U_2=X(\frac{\pi}{2})$ (Figure \ref{fig:180}ii,v), which can be exactly generated using this method.

As shown previously~\cite{luy2005} (Appendix~\ref{app:Luy}), state-to-state excitation ($Z\rightarrow -Y$) and de-excitation ($Y\rightarrow Z$) pulses can be combined to construct unitary pulses that perform $180_x$ (Appendix~\ref{app:Luy}), such as:
\begin{equation}
\begin{split}
V_{180x} &= V_{Y\rightarrow Z} V^\prime_{Y\rightarrow Z}\\
&= V^\prime_{Z\rightarrow -Y} V_{Z\rightarrow -Y}\\
    \end{split}
\end{equation}
whose schematic form will be (Equation \ref{eq:schemSS}, Appendix~\ref{app:Luy})
\begin{equation}
V_{180x} = Z(a\Omega T) X(\pi) Z(a\Omega T).
\end{equation}
As the $a$ value of the parent state-to-state pulses vary with $\Omega$, such a pulse will not be band-schematic and will have offset dependent scalar coupling evolution times. Similarly, the \verb|Q3| pulse~\cite{emsley1992} well performs the action of $180_x$ overall, but is not band-schematic, and so care must be used when using it for refocusing in multi-spin sequences to avoid a variation in evolution time versus $\Omega$. GRAPE inversion pulses and adiabatic inversions also do not control the pre- and post- evolution periods, but they can be relatively short in duration which can mitigate this (Table \ref{tab1}, Figure \ref{fig:180}). In a general case, pulses can be constructed for any purpose described here using Seedless~\cite{seedless}.

\subsubsection{Evolution controlled identity pulses}
\label{sec:ident}
A pulse designed to have the effect of the identity operator will lead to $V_\mathrm{S}$ being independent of $a-b$, as the pre- and post- evolution periods can simply be combined. This can in principle lead to unwanted evolution under scalar coupling. Should there be a need, an evolution controlled identity pulse can be created by concatenating two evolution controlled \verb|a180xa| pulses (section \ref{sec:180x}). 

\subsection{Use of evolution controlled pulses in INEPT transfers}
\label{sec:INEPT}
The INEPT sequence performs the transfer $I_z \rightarrow 2I_zS_z$ and its reverse and is widely used in multi-spin pulse sequences. It requires precise control of scalar-coupling evolution while refocusing chemical shift. Free precession (evolution in the absence of RF pulses) for a time $t$ will be indicated in this section by $F(t)$. In this case, the $S$ spin will remain in terms of populations and so $S$  chemical shift can be neglected, with $F$ describing evolution under $I$ chemical shift and $IS$ scalar coupling. The propagator for an INEPT will be 
\begin{equation}
 V_{\mathrm{INEPT}}= \textcolor{blue}{V_{\mathrm{I},90y}} \,\,F(\tau)\,\, \textcolor{blue}{V_{\mathrm{I},180x}} \textcolor{red}{V_{\mathrm{S},180x}}\,\,F(\tau) \,\, \textcolor{blue}{  V_{\mathrm{I},90x}}
\end{equation}
where the central 180$^\circ$ pulses are typically applied simultaneously. The schematic form for pulses on the $I$ spin can be expressed using $F(t)$ for the evolution periods and provided $V_{\mathrm{S},180x}$ is short, we can write 
\begin{equation}
V_{\mathrm{INEPT}}= \textcolor{blue}{F(\beta_3)Y_\mathrm{I}(\frac{\pi}{2})F(\alpha_3) }\,\,F(\tau)\,\, \textcolor{blue}{F(\alpha_2) X_\mathrm{I}(\pi)}\textcolor{red}{V_{\mathrm{S},180x}  }\textcolor{blue}{  F(\alpha_2)} \,\,F(\tau) \,\,\textcolor{blue} {F(\beta_1)X_\mathrm{I}(\frac{\pi}{2})F(\alpha_1)}.
\end{equation}
The active time for scalar coupling will be $\Delta = \beta_1+2 \tau + 2 \alpha_2 + \alpha_3$, where maximum transfer occurs when  $\Delta = \frac{1}{2J}$. Evolution under $I$ chemical shift will be $\beta_1 - \alpha_3$, which we want to be zero. There are several ways to accomplish this. With rectangular pulses near resonance, we incur additional evolution durations $\alpha=\beta=\frac{2}{\pi}T_{90}$, and
\begin{equation}
 \Delta_{\mathrm{HARD}} = 2\tau + 4 \frac{2}{\pi}T_{90}  
\end{equation}
where $T_{90}$ is the duration of the 90$^\circ$ rectangular pulses~\cite{kay1990}. For maximum transfer we require $\Delta=\frac{1}{2J}$. These pulses are not robust to variation in $B_1$ and have limited bandwidth.

\begin{figure}
     \includegraphics[trim={2cm 21cm 2cm 2cm},width=1.05\linewidth]{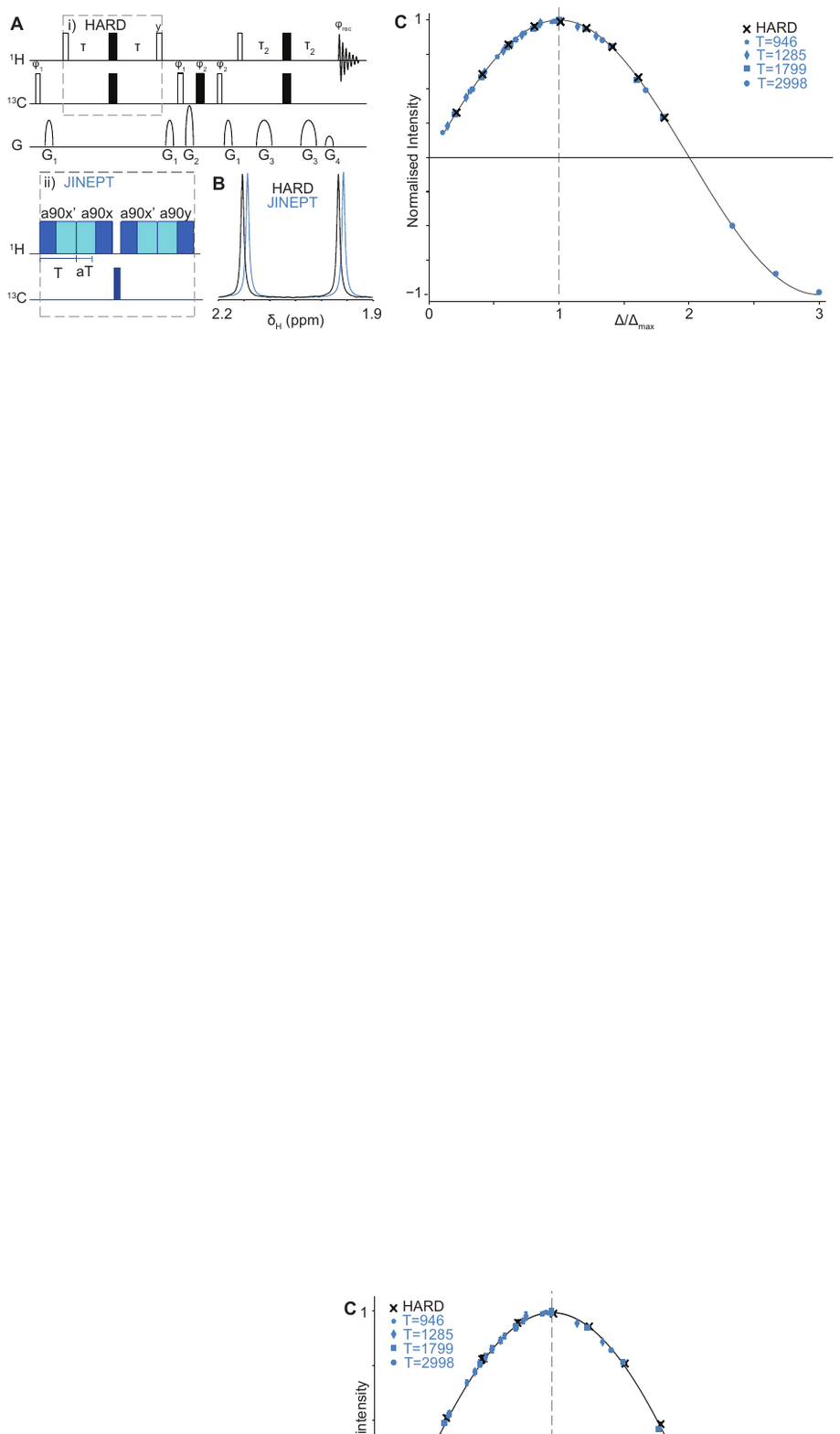}
     \caption{\textbf{A} A 1D HSQC experiment was adapted so that the first INEPT consisted either of rectangular pulses (HARD) or a continuously pulsed JINEPT element (Equation \ref{eq:JINEPT}). The time that scalar coupling time is active $\Delta$ could be varied by changing $\tau$ (HARD) or by altering $aT$ in the a90x pulses (JINEPT).
     Pulse phases are on $x$ unless otherwise indicated with $\phi_1 = (0,2)$, $\phi_2=(0,0,2,2)$ and $\phi_{\text{rec}}=(2,0,0,2)$. Spectra were recorded with $\tau_2=1.75$ ms and gradients 1-4 were applied with durations (1, 1, 1, 0.15) ms at (20, 80, 5, 20.1)\% of the maximum. Rectangular pulses were calibrated to have a $90^\circ$ rotation time of 7.84 and 21.46 $\mu$s on $^1$H and $^{13}$C respectively at maximum power. They were applied on-resonance to maximise their performance, which will deteriorate moving off-resonance~\cite{seedless}. 41 different JINEPT sequences were tested using a90x seedless pulses with varying $T$ and $a$. These were all applied at a field equal to half of the maximum possible (15.7 KHz, see Appendix \ref{app:expmeth}), ensuring that the probe was not overloaded. A rectangular 180$^\circ$ pulse was used on the $^{13}$C in the JINEPT element. Switching all pulses to GRAPE pulses is expected to boost overall sensitivity by ca. 30\% at 750 MHz~\cite{seedless}.    
     \textbf{B}
     Spectra of $^{13}$CH$_3$ methyl labelled methionine were acquired, revealing a doublet. Both HARD and JINEPT spectra are shown with $\Delta=\Delta_{max}=\frac{1}{2J}$ leading to maximum sensitivity (the JINEPT spectrum has been offset for clarity). \textbf{C} Peak intensities of both spectra varied with $\Delta$. For HARD, $\tau$ was varied. For JINEPT, four total durations ($T =$ 946, 1285, 1799 and 2998 $\mu$s) were tested with $a$ ranging from 0.1 to 0.95. The datasets were combined and fitted to $I=A\sin(\Delta J\pi)$, where the fitted value of $J$ was 138.6 Hz and $\Delta_{max}$ =3.6 ms. This demonstrates that varying $\tau$ (HARD) and adjusting $aT$ (JINEPT) leads to equivalent variations in signal intensity, and that scalar coupling can be  controlled with these pulses. All spectra are shown in Figure \ref{fig:processed_spectra}.}
      \label{fig:JINEPT}
\end{figure}

It can be beneficial to use state-to-state pulses. To minimise its duration, $V_{\mathrm{S},180x}$  can be a $Z\rightarrow -Z$ inversion. Typically, we require a pulse with symmetric frequency response about the carrier which is guaranteed by imposing time reflection symmetry (Appendix~\ref{app:timePhaseRev}). This can be accomplished in a Seedless pulse calculation by specifying \verb|SYM| in the input script (Appendx~\ref{app:SeedScript}). Short phase-only inversion pulses computed in Seedless tend to have a smooth profile and resemble the widely used BIP pulses~\cite{smith2001} (Figure \ref{fig:180}iii). Because the pre- and post- evolution delays are unequal for an inversion pulse, $J$ will be active for a time $(a-b)T$, and $I$ spin chemical shift will evolve for a duration $T$. Both effects will introduce errors, and so the main requirement is that this pulse is short, a requirement that can be easily accommodated using Seedless (Figure \ref{fig:180}iii).

A state-to-state ($Z\rightarrow-Y$) pulse is well suited for $I$ spin excitation, and its order/phase reversed symmetry partner can be used for de-excitation ($Y\rightarrow Z$)~\cite{seedless} to exactly balance the evolution periods. In the schematic picture above, these set $\alpha_3=\beta_1=0$ (which are band-schematic), and $\alpha_1=\beta_3$ (which are not). As magnetisation is longitudinal during the periods described by $\alpha_1$ and $\beta_3$, it is not important in this application that these two values are uncontrolled. Taken together, $\tau=\frac{1}{2J}-2\alpha_2$. 

To control the evolution delays $\alpha_2$, the central refocusing $I$ pulse can be constructed from two \verb|a90x| pulses (section \ref{sec:180x}). These approaches collectively allow rectangular pulses to be exchanged for evolution controlled GRAPE pulses to expand effective bandwidth and to compensate for $B_1$ inhomogeneity.

We illustrate these concepts with the joint INEPT (JINEPT), in which the $I$ spin is irradiated continuously using evolution-controlled $90^\circ$ pulses. In this sequence, free evolution delays are removed entirely and the total scalar-coupling evolution is only set by the effective evolution parameters of the pulses. This requires computation of one \verb|a90x| pulse of duration $T$ (an evolution controlled $90^\circ$ pulse with $b=0$) acting on the $I$ spin and its order/phase reversed partner. We then perform
\begin{equation}
V_{\mathrm{JINEPT}}=\textcolor{blue}{V_{\mathrm{I},a90y}}  \,\,\, \textcolor{blue}{V^\prime_{\mathrm{I},a90x}} \,\,\, \textcolor{red}{V_{\mathrm{S},180x}} \,\,\, \textcolor{blue}{V_{\mathrm{I},a90x}} \,\,\, \textcolor{blue}{V^\prime_{\mathrm{I},a90x}}.
  \label{eq:JINEPT}
    \end{equation}
The $V_{\mathrm{S},180x}$ should be brief. The phase shift to $y$ on the final pulse corresponds to rotating the original pulse rotated about the $z$ axis by $90^\circ$, accomplished either by adjusting the phase from 0 to 1 in the pulse sequence, or adding $\pi/2$ to all the phase of all elements. The schematic propagator is
\begin{equation}
V_{\mathrm{JINEPT}}=\textcolor{blue}{Y_\mathrm{I}(\frac{\pi}{2}) F(\alpha)}  \textcolor{blue}{ F(\alpha) X_\mathrm{I}(\frac{\pi}{2}) }\textcolor{red}{V_{\mathrm{S},180x}} \textcolor{blue}{ X_\mathrm{I}(\frac{\pi}{2}) F(\alpha) } \textcolor{blue}{ F(\alpha) X_\mathrm{I}(\frac{\pi}{2})}
    \end{equation}
leading to
\begin{equation}
    \Delta_{\mathrm{JINEPT}} = 4aT.
\end{equation}
Experimental validation of the JINEPT was performed using a modified HSQC experiment (Figure \ref{fig:JINEPT}) where the first INEPT transfer was either performed using rectangular pulses applied on-resonance to maximise their performance (HARD, Figure \ref{fig:JINEPT}A) or a JINEPT. Spectra of $^{13}$CH$_3$-labelled methionine were acquired, and the final intensity was monitored while varying $\Delta$ (Figure \ref{fig:JINEPT}C). The intensities lie on the curve $A\sin(\Delta J\pi)$ revealing that we obtain identical behaviour by either controlling the scalar coupling by altering a delay, or by altering the period in which scalar coupling is allowed during an evolution controlled pulse.

The total duration of the INEPT element using rectangular pulses (including pulses) was 3,599 $\mu$s which will be close to the shortest total duration possible. The total duration of the JINEPT with $T=946$ $\mu$s and $a=0.95$ was 3,784 $\mu$s, resulting in a relatively modest increase in the total duration of 185 $\mu$s. Retaining the delay and using shorter GRAPE pulses at higher powers will allow INEPT elements to be constructed of intermediate duration. These results confirm that evolution-controlled pulses designed using single-spin optimisation can accurately control ensembles of weakly coupled spins while maintaining band selectivity and $B_1$ robustness (Figure \ref{fig:JINEPT}). 

\section{Discussion} 
RF pulses optimised for an isolated spin-1/2 can be created using GRAPE methods to precisely control the evolution of ensembles of weakly coupled spins. This requires that only one spin is excited at any time, and that the pulse is band-schematic. The pulsed spin's chemical shift and scalar couplings will evolve according to the schematic form, with pre- ($aT$) and post- ($bT$) delays and all other chemical shifts and couplings will evolve for a duration $T$. This works for arbitrarily sized spin systems, including nuclei of any spin.

Evolution-controlled pulses extend the utility of single-spin optimal control to multi-spin experiments without the computational burden of full multi-spin optimisation. The approach naturally accommodates band selectivity and $B_1$ robustness and is particularly advantageous in sequences, such as INEPT, where accurate timing of scalar-coupling evolution is essential. The JINEPT sequence illustrates how continuous irradiation with evolution-controlled pulses can eliminate free delays while preserving the desired transfer pathway.

The framework breaks down in regimes involving strong coupling or simultaneous pulsing on multiple spins. Here, true multi-spin optimisation or sufficiently short pulses (where $T<<\frac{1}{J}$) are required. Moreover, we neglect relaxation effects. Within its domain of validity, however, the approach provides a practical and transparent route to robust multi-spin control.

These methods and analysis options have been implemented in the software Seedless, which can be freely downloaded. Computing the pulses for common biomolecular and chemical NMR applications takes less than a second on a M4 pro mac laptop and so can be generated `on-the-fly', to best match a given sample/hardware combination of current interest.

\section{Download}
Seedless is free for academic use and can be downloaded from 
\verb|https://seedless.chem.ox.ac.uk|. The manual provided describes the syntax required for the input in detail, and examples of pulses described in this manuscript are provided as demonstrations. 

\section{Acknowledgements}
We are grateful to Lewis Kay for many things, but here, for useful discussions and confirming that he subtracted $\frac{2}{\pi}$T from the rectangular pulses in the original implementations of triple resonance experiments following advice from Ad Bax. We are also grateful for discussions with the creator of the GRAPE method, Steffen Glaser, up a mountain in Switzerland who has previously explored many of these ideas. We thank Philip Wurm, Maksim Mayzel and Niels Karschin (Bruker) for helpful discussions and noting Seedless GRAPE inversion pulses converge on BIP pulses. AJB has received funding from the European Research Council (ERC) under the European Union’s Horizon 2020 research and innovation programme (grant agreement No 101002859)

\bibliography{pulsesBib}

\appendix

\section{One-spin methods and block diagonal matrices}
\label{isomorph}
We wish to understand when and how multi-spin systems can be controlled using concepts designed for the treatment of isolated spin-\half\ nuclei. For simplicity we start with the case of two weakly-coupled spin-\half\ nuclei, for which the background Hamiltonian takes the form
\begin{equation}
\mathcal{H}_0=\Omega_I I_z +\Omega_S S_z+J I_z S_z.
\end{equation}
We will show later that equivalent results apply to an number of spins of any type, as long as high-spin nuclei can be described solely by linear operators, either because quadrupolar and higher terms are naturally zero or because they have been motionally averaged to zero. The restriction to weak coupling is, however, important.

We are interested in the effect of applying excitations to one of the two spins. For our two-spin system we will pulse spin $S$ along the $x$-axis in the rotating frame, so that
\begin{equation}
\mathcal{H}_1=\omega S_x. 
\end{equation}
The choice of spin $S$ means that arguments based on the block-diagonal structure of matrix representations will be particularly obvious, but we can switch $S$ for $I$ and make an entirely equivalent argument, as long as only one spin (or more generally only one of a group of spins) is excited. Equivalent arguments will also work for any excitation phase. 

Since the background and excitation Hamiltonians do not commute it might appear necessary to perform all calculations using the total Hamiltonian $\mathcal{H}=\mathcal{H}_0+\mathcal{H}_1$, but this is not the case. First notice that the term $\Omega_I I_z$ commutes with the rest of $\mathcal{H}_0$ and with $\mathcal{H}_1$ and so the evolution under this term can be calculated separately as
\begin{equation}
V=\rme^{-\rmi\mathcal{H}t}=\rme^{-\rmi\Omega_I I_z t}\rme^{-\rmi\mathcal{H'}t}
\end{equation}
where
\begin{equation}
\mathcal{H'}=\Omega_S S_z+J I_z S_z+\omega S_x
\end{equation}
is the remaining Hamiltonian, including the excitation term and everything which does not commute with it. For a multispin system all the Zeeman terms on unexcited spins, and all the couplings between such spins, can be treated in this way. Writing out $\mathcal{H'}$ in explicit matrix form
\begin{equation}
\mathcal{H'}=\begin{pmatrix}
\half\Omega_S+\quarter J&\half\omega&0&0\\
\half\omega&-\half\Omega_S-\quarter J&0&0\\
0&0&\half\Omega_S-\quarter J&\half\omega\\
0&0&\half\omega&-\half\Omega_S+\quarter J
\end{pmatrix}
\end{equation}
immediately reveals that it has a block-diagonal form
\begin{equation}
\mathcal{H'}=\begin{pmatrix}
\mathcal{H}_+&\mathbf{0}\\
\mathbf{0}&\mathcal{H}_-
\end{pmatrix}
\end{equation}
where the two $2\times2$ blocks
\begin{equation}
\mathcal{H}_\pm=\begin{pmatrix}
\half(\Omega_S\pm\half J)&\half\omega\\
\half\omega&-\half(\Omega_S\pm\half J)
\end{pmatrix}=\begin{pmatrix}
\half\Omega_\pm&\half\omega\\
\half\omega&-\half\Omega_\pm
\end{pmatrix}
\end{equation}
have exactly the same form as a one-spin Hamiltonian describing excitation along $x$ in the presence of an offset $\Omega_\pm=\Omega_S\pm\half J$. Since block diagonal structure is preserved under matrix additions and multiplications, the structure in the Hamiltonian leads to an identical block diagonal structure in the propagator for the pulse. Thus the total evolution under $\mathcal{H}$ can be evaluated by combining evolution under $\Omega_I I_z$ for a time $t$ with evolution under the remainder of the Hamiltonian calculated by replacing it with an equivalent pair of one-spin Hamiltonians. In the same way the total propagator for a sequence of pulses, or for a shaped pulse, will have the same block diagonal structure, and can be worked out using one spin methods in an equivalent way. Thus we can write
\begin{equation}
V=\rme^{-\rmi\Omega_I I_z t}\begin{pmatrix}
V_+&\mathbf{0}\\
\mathbf{0}&V_-
\end{pmatrix}
\end{equation}
where $V_\pm$ are one-spin propagators.

Note that nothing in this argument requires that there be only two blocks, and thus an equivalent approach can be used for any number of spins as long as only one of them is excited.  All the terms which commute with $S_x$ will evolve for the full time $t$, while all the non-commuting terms are incorporated into $\mathcal{H'}$, which now contains a larger number of one-spin blocks. The same thing remains true if the unexcited nuclei are high-spin (spin greater than $\frac{1}{2}$): once again this simply increases the number of separate blocks, each of which continues to have a one-spin form with its own offset frequency. Finally, if the excited spin is high-spin then the block diagonal structure will remain, but the blocks will no longer be $2\times2$, instead having a structure corresponding to a single high-spin nucleus with suitable offset frequencies. For the purposes of GRAPE pulse computation, if the high-spin nucleus has no quadrupolar or higher interactions then it will evolve in a manner that is isomorphic to a spin-\half\ (the density matrix and Hamiltonians are written with the same spin operators as spin-\half, which have identical commutators), and so can be treated as if it were spin-\half\ for our purposes.

Until this point every step in the argument has been exact, but we now add a single further assumption, which will only be approximately true. Suppose that the two single-spin evolutions are well described by the same schematic form, so that
\begin{equation}
V_\pm\approx\rme^{-\rmi b \Omega_\pm A_z t}U\rme^{-\rmi a \Omega_\pm A_z t}
\end{equation}
where $A_z=\half\sigma_z$ is a one-spin Zeeman operator, which will be true if the one-spin evolution is well described by a band-schematic pulse, with both values $\Omega_\pm=\Omega_S\pm\half J$ lying within the schematic band. In this case the whole evolution can be described by a schematic form, with evolution under combined $\Omega_S$ and coupling for a time $a t$, followed by the central fixed unitary, and then further evolution under combined $\Omega_S$ and coupling for a time $b t$. Finally the evolution under $\Omega_I$ (and all other commuting evolutions) for time $t$ (the whole length of the pulse) can be applied at any stage in the calculation. The treatment of $\Omega_I$ evolution is exact, as long as excitation is confined to spin $S$.

Note that this argument constructs an approximate propagator describing evolution under $\mathcal{H'}$, which will be a good approximation provided that the band schematic conditions are met. As we are constructing a propagator this will apply to any initial state, including multiple quantum coherences. The only restriction is that the excitation field is only ever applied to one spin at a time. In cases where spins other than the one being excited are transverse, care will be needed to properly account for the evolution, where it will typically be desirable to have the pulses simply be short enough that they can be treated as an instantaneous rotation.

\section{Calculation of schematic pulse parameters}
\label{app:opdev}
For a pulse whose propagator is known at a given offset, we want to obtain the schematic pulse parameters. For the special case of a rectangular pulse with a small flip angle ($\theta=\omega t$) close to resonance, that is with both $\theta $ and $\Omega t$ much less than one, the schematic form can be deduced using the Trotter--Suzuki formula
\begin{equation}
V=\rme^{-\rmi(\omega I_x+\Omega I_z)t}\approx\rme^{-\rmi\Omega I_z t/2}\rme^{-\rmi \omega I_x t}\rme^{-\rmi \Omega I_z t/2},
\end{equation}
that is the pulse can be replaced by a zero-time rotation with the same flip angle preceded and followed by evolution periods of length $t/2$, and so $a=b=\frac{1}{2}$. By comparing the fidelity of the schematic pulse to its exact form and expanding to first order in $\Omega$, we can see that for larger flip angles the same general form remains, but now $a=\tan(\theta/2)/\theta$, so that for the special case of a $90^\circ$ rotation $a=2/\pi$. This approach breaks down for rectangular pulses far from resonance or, particularly relevant for this work,  shaped pulses including phase and amplitude modulation.

\subsection{Derivation of evolution operator $p$}
For a general propagator $V=\rme^{-\rmi t \mathcal{H}}$, we will instead first perform a one-sided decomposition, $V=UP$ where $P=\rme^{-\rmi t p \epsilon}$ where $\epsilon$ is a scalar. Taking a lead from Feynman and Waugh, starting from the general derivative of an exponential map~\cite{Feyn1951,wilcox1967,waugh1968},
\begin{equation}
\frac{\rmd}{\rmd\epsilon}e^X = e^X \int_0^1 e^{-sX}\frac{\rmd X}{\rmd\epsilon}e^{sX} \rmd s,
\label{eq:ExpMap}
\end{equation}
we can write the dependence on $\Omega$ as
\begin{equation}
\frac{\rmd V}{\rmd \Omega}= -\rmi t V \int_0^1 \rme^{s(\rmi t\mathcal{H})}\frac{\rmd  \mathcal{H}}{\rmd\Omega}e^{-s(\rmi t \mathcal{H})} \rmd s.
\end{equation}
Substituting $st=t^\prime$ then 
\begin{equation}
\frac{\rmd V}{\rmd \Omega}= -\rmi t V \frac{1}{t}\int_0^t \rme^{\rmi t^\prime \mathcal{H}}\frac{\rmd  \mathcal{H}}{\rmd\Omega}e^{-\rmi t^\prime \mathcal{H}} \rmd t^\prime.
\end{equation}
As $V^\dagger\frac{\rmd V}{\rmd\Omega}$ is anti-Hermitian, it can be associated with a rotation $P=\rme^{V^\dagger\frac{\rmd V}{d\Omega}\Omega}$, allowing us to identify a dimensionless Hermitian matrix
\begin{equation}
p=\frac{\rmi}{t}V^\dagger\frac{\rmd V}{\rmd \Omega}=  \frac{1}{t}\int_0^t \rme^{\rmi t^\prime \mathcal{H}}\frac{\rmd  \mathcal{H}}{\rmd\Omega}e^{-\rmi t^\prime \mathcal{H}} \rmd t^\prime,
\label{eq:AveHam}
\end{equation}
which is Equation \ref{eq:pdef}. By defining the propagator $P=\rme^{V^\dagger\frac{\rmd V}{\rmd \Omega}\Omega}=\rme^{-\rmi t p \Omega}$, we complete the decomposition $V=UP$, which is exact. The definition for $p$ closely resembles the first order average Hamiltonian widely used in NMR. In this case $\frac{\rmd  \mathcal{H}}{\rmd\Omega}$ isn't a small perturbation, and all other terms in the Magnus expansion are effectively contained within $U$. In this work we use $\frac{\rmd \mathcal{H}}{\rmd \Omega}=I_z$. 

Next we explore the derivative of $U$, 
\begin{equation}
\begin{split}
\frac{\rmd U}{\rmd\Omega} &= \frac{\rmd (VP^\dagger)}{\rmd\Omega}= \frac{\rmd V}{\rmd\Omega}P^\dagger + V \frac{\rmd P^\dagger}{\rmd\Omega}\\
\frac{\rmd U}{\rmd\Omega} &= -\rmi t Vp P^\dagger + V \frac{\rmd P^\dagger}{\rmd\Omega}.\\
\end{split}
\end{equation}

Noting $P^\dagger= e^{+i t p \Omega}$ then we can re-use Equation \ref{eq:ExpMap} and
\begin{equation}
\begin{split}
\frac{\rmd U}{\rmd \Omega} &= -\rmi t Vp P^\dagger + \rmi t V (P^\dagger  \frac{1}{t}\int_0^t \rme^{-\rmi t^\prime p \Omega}\frac{\rmd  (p \Omega)}{\rmd\Omega}  e^{\rmi t^\prime p \Omega}\rmd t^\prime )\\
 &= -\rmi t Vp P^\dagger + \rmi t V (P^\dagger  \frac{1}{t}\int_0^t e^{-\rmi t^\prime p \Omega}( p + \frac{\rmd p}{\rmd\Omega}\Omega )  e^{\rmi t^\prime p \Omega}\rmd t^\prime )\\
\end{split}
\end{equation}
As $p$ commutes with $P$ we can simplify to
\begin{equation}
\begin{split}
\frac{\rmd U}{\rmd\Omega}  &=\rmi  \Omega U \int_{0}^t e^{-\rmi t^\prime p \Omega}\frac{\rmd p}{\rmd\Omega}e^{\rmi t^\prime p\Omega} \rmd t^\prime .
 \label{eq:Uderiv}
\end{split}
\end{equation}
For the pulse to have the same schematic parameters over a range of evolution frequencies and be `band-schematic' in this region requires  $\frac{\rmd p}{\rmd\Omega}=0$, from which it follows that $\frac{\rmd U}{\rmd\Omega}=0 $, as stated in the text. 

\subsection{Calculation of $p$}
\label{app:calcP}
We first compute the evolution matrix $p$ (Equation \ref{eq:pdef}) for a rectangular pulse acting on a single spin-\half\ where $\mathcal{H} = \omega(\cos\phi I_x + \sin\phi I_y) + \Omega I_z$. It is convenient to project $p$ onto a Cartesian basis $I=(I_x,I_y,I_z)$ to obtain a fractional `evolution vector', $\Vec{p}=(p_x,\,p_y,\,p_z)$ where $p=\Vec{p}\cdot I$. The result is cylindrically symmetric
\begin{equation}
p_x=p_x^0 \cos\phi - p_y^0\sin\phi,\quad p_y=p_x^0\sin\phi + p_y^0\cos\phi,\quad p_z=p_z^0,
\end{equation}
where $\psi=t\sqrt{\omega^2+\Omega^2}$ is the rotational angle around the tilted rotation axis $\frac{t}{\psi}(\omega\cos\phi,\omega\sin\phi,\Omega)$ and
\begin{equation}
p_x^0=\frac{t^2 \omega \Omega}{\psi^2} \left(1-\mathrm{sinc }\, \psi\right)  ,\quad p_y^0=  \frac{t\omega }{\psi^2}\left(1-\cos \psi\right),\quad p_z^0=\frac{t^2}{\psi^2}\left(\Omega^2+\omega^2\mathrm{sinc}\, \psi\right).
\end{equation}

We need to analyse shaped pulses. Here, the propagator is comprised of a product of $N$ individual elements of duration $t$ giving a total duration $T$,
\begin{equation}
V=V_N V_{N-1}\dots V_3 V_2 V_1 = \prod_{j=N}^1 V_j.
\label{eq:bigV}
\end{equation}
Using the definition of $p$ (Equation \ref{eq:pdef}),
\begin{equation}
\begin{split}
p=\frac{\rmi}{T}V^\dagger \frac{d V}{d\Omega}&=\frac{\rmi}{T}V^\dagger 
\sum_{k=1}^N \left(\prod_{n=N}^{k+1} V_n\right) \frac{d V_k}{d\Omega} \left(\prod_{n=k-1}^{1} V_n\right) ,
 \end{split}
\end{equation}
then defining the partial products $X_j=V_jX_{j-1}=V_j V_{j-1}...V_2 V_1= \prod_{n=j}^1 V_n $, and $V X^\dagger_{j-1} = \prod_{n=N}^j V_n$ (where $X_0$ is the identity),
\begin{equation}
\begin{split}
p=  &=\frac{\rmi}{T}\sum_{k=1}^N  V^\dagger V X_k^\dagger \frac{d V_k}{d\Omega} X_{k-1} \\
&=\frac{\rmi}{T}\sum_{k=1}^N  X_k^\dagger \frac{d V_k}{d\Omega} X_{k-1} .\\
\end{split}
\end{equation}
For each element, we can write $\frac{d V_k}{d\Omega}=\frac{t}{\rmi} V_k p_k$ (Equation \ref{eq:pdef}). Noting that $\frac{t}{T}=\frac{1}{N}$ we obtain 
\begin{equation}
p= \frac{1}{N}\sum_{k=1}^N  X_{k-1}^\dagger p_k X_{k-1}.
\label{eq:pshape}
\end{equation}
This is the average of the individual $p_k$ values, with each transformed into an interaction frame equivalent to the action of the pulse up to each element. If all the elements in the shaped pulse are identical, we can take the continuous limit and recover Equation \ref{eq:AveHam}. Thus $p$ can be computed for any offset for an arbitrary pulse, and used with $V$ to get the local schematic pulse parameters using Equation \ref{eq:decomp}. 

In Seedless~\cite{seedless} this computation is performed as part of final analysis and not during the optimisation routines where it isn't efficient to explicitly compute the partial products. Instead, we set all elements in the matrix $Q_{N+1}$ to zero and recursively compute $Q_k=V_k^\dagger Q_{k+1} V_k + p_k$ working backwards from $k=N$ to $k=1$ where $p=\frac{1}{N}Q_1$. When \verb|EVOLVE| is specified, the values $a$, $b$, and the axis/angle of $U$ vary for a pulse of interest are plotted as a function of $\Omega$ for any pulse either created, or read in, for analysis (e.g. Figures \ref{fig:90} and \ref{fig:180}). If the parameters are constant over a range of frequencies, the pulse can be considered band-schematic over this region and, for example, the durations $aT$ and $bT$ can be subtracted from any delays flanking the pulse.

\subsection{Calculation of schematic pulse parameters}
\label{app:calcSchem}
The propagator $V$ (Equation \ref{eq:bigV}) and evolution vector $\Vec{p}$ (Equation \ref{eq:pshape}) can be computed for any pulse at any offset. We seek a method to calculate the schematic pulse parameters for a shaped pulse, $U$, $a$ and $b$ from these. Describing the central rotation in $V_\mathrm{S}$ (Equation \ref{schem}) by a unit vector $n=(\sin u_\theta \cos u_\phi, \sin u_\theta \sin u_\phi, \cos u_\theta)$ and angle $u_\psi$, then the matrix form of the schematic pulse  is
\begin{equation}
V_{\mathrm{S}}=
\begin{pmatrix}
 \rme^{-\frac{1}{2} \rmi (a+b)\Omega t} \left(\cos \frac{u_\psi}{2}-\rmi  \cos u_\theta \sin \frac{\psi}{2}\right) 
 & -\rmi \rme^{\frac{1}{2} \rmi (a-b)\Omega t-\rmi u_\phi}  \sin\frac{u_\psi}{2}\sin u_\theta \\
 -\rmi \rme^{-\frac{1}{2} \rmi (a-b)\Omega t + \rmi u_\phi}  \sin \frac{u_\psi}{2} \sin u_\theta 
 & \rme^{\frac{1}{2} \rmi (a+b)\Omega t} \left(\cos \frac{u_\psi}{2}+i \cos u_\theta \sin \frac{u_\psi}{2}\right) \\
\end{pmatrix}.
\label{eqUni}
\end{equation}
We need the schematic pulse parameters that lead to $V\equiv V_\mathrm{S}$. Projecting both onto the basis $I=(I_x,I_y,I_z,E)$, then
\begin{equation}
\left(
\begin{array}{c}
V_x \\ V_y \\ V_z \\ V_E
\end{array}
\right)
\equiv\left(
\begin{array}{c}
-2 i \sin \left(\frac{u_\psi }{2}\right) \sin \left(u_\theta \right) \cos \left(\frac{1}{2} \Omega t  (a -b )-u_\phi \right)\\
2 i \sin \left(\frac{u_\psi }{2}\right) \sin \left(u_\theta \right) \sin \left(\frac{1}{2} \Omega t  (a - b )-u_\phi \right)\\
-2 i \left(\cos \left(\frac{u_\psi }{2}\right) \sin \left(\frac{1}{2} \Omega t (a +b )\right)+\sin \left(\frac{u_\psi }{2}\right) \cos \left(u_\theta \right) \cos \left(\frac{1}{2} \Omega t  (a +b )\right)\right)\\
\cos \left(\frac{u_\psi }{2}\right) \cos \left(\frac{1}{2} \Omega  t (a + b )\right)-\sin \left(\frac{u_\psi }{2}\right) \cos \left(u_\theta\right) \sin \left(\frac{1}{2} \Omega t  (a +b )\right)
\end{array}
\right).
\end{equation}
We cannot independently extract the 5 schematic pulse parameters from the 4 projected values of $V$. As $V$ is unitary and $V_x^2+V_y^2+V_z^2=4(V_E^2-1)$, it contains only 3 independent parameters. Given the choice of basis and structure of $V$, the projections $V_x,V_y,V_z$ are imaginary and $V_E$ is real. Because of the cylindrical symmetry it is helpful to define
\begin{equation}
\begin{array}{rl}
V_R^2&=V_x^2+V_y^2=-4 \sin ^2\left(\frac{u_\psi }{2}\right) \sin ^2\left(u_\theta \right). \\
\end{array}
\end{equation}
We can compute a schematic evolution matrix from $V_\mathrm{S}$, $p_\mathrm{S}=\frac{\rmi}{T} V_\mathrm{S}^\dagger \frac{dV_\mathrm{S}}{d\Omega}$, project it onto a Cartesian basis and equate this to the components of $\Vec{p}$
\begin{equation}
\left(
\begin{array}{c}
p_x \\ p_y \\ p_z 
\end{array}
\right)
\equiv
\left(
\begin{array}{c}
 b \left(\sin (u_\psi ) \sin \left(u_\theta \right) \sin \left( a  \Omega t - u_\phi \right)+\sin ^2\left(\frac{u_\psi }{2}\right) \sin \left(2 u_\theta \right) \cos \left(a  \Omega t -u_\phi\right)\right)\\
 b  \left(\sin (u_\psi ) \sin \left(u_\theta \right) \cos \left(a  \Omega t -u_\phi \right)-\sin ^2\left(\frac{u_\psi }{2}\right) \sin \left(2 u_\theta \right) \sin \left(a  \Omega t -u_\phi\right)\right)\\
 a + b\frac{1}{2}  \left(2+ V_R^2\right)\\
\end{array}
\right).
\end{equation}
We now have 6 values from which we can safely extract the 5 schematic parameters. The squared modulus of $\Vec{p}$ is $|p|^2=p_x^2+p_y^2+p_z^2 =  (a+b)^2 +a b V_R^2$. Noting the cylindrical symmetry of $p$
\begin{equation}
\begin{array}{rl}
p_R^2&=p_x^2+p_y^2= -b ^2 V_R^2 \left(1+\frac{V_R^2}{4}\right)
\end{array}
\end{equation}
and we can rearrange to obtain the post-evolution delay (taking the positive solution)
\begin{equation}
b=\frac{2 p_R}{\sqrt{-V_R^2\left(4+V_R^2\right)}}
\end{equation}
which we combine with $p_z$ to obtain the pre-evolution delay
\begin{equation}
a = p_z - \frac{1}{2} b  \left(  2+ V_R^2\right)=  p_z - p_R\frac{ 2+ V_R^2}{\sqrt{-V_R^2\left(4+V_R^2\right)}}. \\
\end{equation}
With $a$ and $b$ we can recover the other values from $V$. We can get $u_\phi$ from
\begin{equation}
\tan\left(\frac{1}{2}\Omega t(a-b)-u_\phi\right)=\frac{V_y}{-V_x}
\end{equation}
and because
\begin{equation}
\begin{array}{rl}
  V_E +  \frac{1}{2}V_z &= e^{- \rmi\frac{a+b}{2}\Omega t} \left(\cos \left(\frac{u_\psi }{2}\right)- \rmi \sin \left(\frac{u_\psi }{2}\right) \cos \left(u_\theta\right)\right), \\
\end{array}
\end{equation}
we obtain $\psi$ from
\begin{equation}
\begin{array}{rl}
\mathrm{Re}\left(e^{\rmi\frac{a+b}{2}\Omega t}\left(V_E+\frac{1}{2}V_z\right)\right)&=\cos\frac{u_\psi}{2}.\\
\end{array}
\end{equation}
Finally we obtain $u_\theta$
\begin{equation}
\begin{array}{rl}
-\frac{1}{\sin\frac{u_\psi}{2}}\mathrm{Im}\left(e^{\rmi\frac{a+b}{2}\Omega t}\left(V_E+\frac{1}{2}V_z\right)\right)&=\cos u_\theta.\\
\end{array}
\end{equation}
Thus for any pair of $V$ and $p$ computed for a given $\Omega$, we obtain their projections and compute the 5 schematic parameters. In summary,
\begin{equation}
\begin{array}{rl}
b&=2 p_R \left(-V_R^2\left(4+V_R^2\right)\right)^{-1/2}\\
a&= p_z - \frac{1}{2}b\left(2+V_R^2\right) \\
u_\phi&=\frac{1}{2}\Omega t(a-b)-\arctan\left(\frac{V_y}{-V_x}\right)\\
u_\psi&=2\arccos\left(\mathrm{Re}\left(e^{\rmi\frac{\alpha+\beta}{2}\Omega t}\left(V_E+\frac{1}{2}V_z\right)\right)\right)\\
u_\theta&=\arccos\left(-\frac{1}{\sin\frac{\psi}{2}}\mathrm{Im}\left(e^{\rmi\frac{a+b}{2}\Omega t}\left(V_E+\frac{1}{2}V_z\right)\right)\right)
\label{eq:decomp}
\end{array}
\end{equation}
An interesting special case occurs when the unitary is a 90$^\circ$ in the $xy$ plane ($u_\psi=u_\theta=\frac{\pi}{2} $). Here, $V_R^2=-2$ and $a=p_z$ and $b=p_R$ and so schematic pulse parameters are obtained from direct inspection of the evolution vector. 

There are two singularities where this method breaks down. Firstly, when the central unitary is a refocusing 180$^\circ$ pulse in the xy plane ($u_\theta = \frac{\pi}{2}$, $u_\psi = \pi$ where $a=b$), and both $a$ and $b$ vanish from $p$. We provide a route to creating such a pulse, as discussed, by splitting it into two (section \ref{sec:180x}). Secondly, when $u_\psi$ is an integer multiple of $2\pi$ (including zero) where $a=b$, which vanish from $p$. A method to overcome this challenge is presented.

\section{Order/phase reversed and time symmetric pulses}
\label{app:timePhaseRev}

\begin{table}
\begin{tabular}{ccccccc}
Symbol   &  Action                          &                 & Starting state             &  Final state                 &  order & phase \\
\hline
$V$        &  $V$                               &  $V^{xyz}(\psi)$  & $\rho_s$                     &  $\rho_f$                      & normal & $\phi$ \\
$V^\prime$ &  $Y(\pi)V^\dagger Y^\dagger(\pi)$  &  $V^{x-yz}(\psi)$ & $Y(\pi)\rho_f Y^\dagger(\pi)$ & $Y(\pi) \rho_s Y^\dagger(\pi)$ & invert &$-\phi$ \\
$V^x$      &  $X(\pi)V^\dagger X^\dagger(\pi)$  &  $V^{-xyz}(\psi)$ & $X(\pi)\rho_f X^\dagger(\pi)$ & $X(\pi) \rho_s X^\dagger(\pi)$ & invert & $-\pi-\phi$\\
$V^z$      &  $Z(\pi)V^\dagger Z^\dagger(\pi)$  &  $V^{xy-z}(\psi)$ & $Z(\pi)\rho_f Z^\dagger(\pi)$ & $Z(\pi) \rho_s Z^\dagger(\pi)$ & invert &$\pi+\phi$\\
\end{tabular}
\caption{Given a shaped pulse comprising $N$ elements and propagator $V$, a new symmetry related partner pulse can be created by inverting the order of the elements, and then adjusting the phases of each element as indicated. The resulting pulse will have a propagator whose action is related to the Hermitian conjugate of the original pulse as shown. Given that the original propagator performs the transform $\rho_s\xrightarrow{V}\rho_f$, the action of the symmetry partner pulses is also shown. The action specifically of the order/phase reversed pulse ($V^\prime$) is related to a $Y(\pi)$ rotation as indicated and is perhaps the most convenient to compute.}
\label{tab:sym}
\end{table}
Symmetries present in the propagators of shaped pulses in the time domain can lead to useful properties. These have been considered in detail~\cite{levitt1982,PulseSymNgoMorris1987}. It is helpful to re-derive some of these results in the notation used in this work. The propagator of a shaped pulse is (noting explicitly the $x,y,z$ components of the corresponding Hamiltonian) 
\begin{equation}
V=V_N^{x,y,z}...V_2^{x,y,z}\, V_1^{x,y,z},
\end{equation}
whose effects could be inverted if we can create a pulse whose propagator behaves as its Hermitian conjugate,
\begin{equation}
V^\dagger =V_1^{-x,-y,-z}\, V_2^{-x,-y,-z} ... V_N^{-x,-y,-z}
\end{equation}
We cannot directly construct a pulse that behaves this way as we would need to either invert the sign of the time operator or invert the direction of the static magnetic field during the experiment. We can construct a pulse that comes close to behaving in this way by creating a new pulse from the first where we reverse the order in which we apply the elements, and inverting the sign of the phase of each (order/phase reversal). The phase reversal inverts the sign of the $y$ axis of each element leading to 
\begin{equation}
V^\prime=V_1^{x,-y,z} \, V_2^{x,-y,z} ... V_N^{x,-y,z}
\end{equation}
The Hermitian conjugate of the original pulse and the order/phase reversed pulse are related by
\begin{equation}
V^\prime = Y(\pi) V^\dagger Y^\dagger(\pi).
\end{equation}
If the original pulse has the schematic propagator $V= Z(\beta) U^{x,y,z} Z(\alpha)$, then if we order/phase reverse the original, we get a new pulse with the following schematic form,
\begin{equation}
\begin{split}
V^\prime&=Y(\pi) Z(-\alpha) U^{-x,-y,-z} Z(-\beta) Y^\dagger(\pi) \\
&=Y(\pi) Z(-\alpha)Y^\dagger\,(\pi)Y(\pi) U^{-x,-y,-z} Y^\dagger(\pi)\,Y(\pi) Z(-\beta) Y^\dagger(\pi) \\
&= Z(\alpha) Y(\pi) U^{-x,-y,-z} Y^\dagger (\pi) Z(\beta) \\
&= Z(\alpha)  U^{x,-y,z} Z(\beta). \\
\end{split}
\end{equation}
In the specific case where the central unitary acts in the $xz$ plane then $ Y(\pi) U^{-x,0,-z} Y^\dagger (\pi)=U^{x,0,z}$ and so
\begin{equation}
\begin{split}
V  &= Z(\beta)  U^{x,0,z} Z(\alpha) \\
 V^\prime &= Z(\alpha)  U^{x,0,z}  Z(\beta) \\
\end{split}
\end{equation}
Provided that the central unitary is applied along the xz plane, then the action of order/phase reversing the first pulse is to flip the order of the pre- and post- evolution delays, while retaining the same central rotation.

Next we consider the action of a state-to-state pulse, and how this is affected by order/phase reversal. Let a pulse perform the following state-to-state transform
\begin{equation}
    \rho_f = V \rho_s V^\dagger.
\end{equation}
If we perform a 180$^\circ$ rotation about the Y axis on both sides, and insert an identity element
\begin{equation}
\begin{split}
  Y(\pi)  \rho_f Y^\dagger(\pi)&= Y(\pi) V  Y^\dagger(\pi) Y(\pi) \rho_s  Y^\dagger(\pi) Y(\pi) V^\dagger Y^\dagger(\pi) \\
  &= V^{\prime\dagger} Y(\pi) \rho_s  Y^\dagger(\pi) V^\prime \\
    \end{split}
\end{equation}
then we reveal the propagators for the order/phase reversed version of the pulse. Rearranging we reveal the transform performed by the order/phase reversed pulse
\begin{equation}
Y(\pi) \rho_s  Y^\dagger(\pi) =  V^\prime  Y(\pi)  \rho_f Y^\dagger(\pi) V^{\prime\dagger}\\
\end{equation}
We can summarise the action of the two transformations performed by the pulse and its order/phase reversed partner,
\begin{equation}
    \begin{split}
    \rho_s &\xrightarrow{V} \rho_f \\
    Y\rho_f Y^\dagger &\xrightarrow{V^\prime} Y \rho_s Y^\dagger \\
    \end{split}
\end{equation}
To take an example, if the original pulse is for excitation, $I_z\xrightarrow{V}-I_y$ then $-I_y \xrightarrow{V^\prime} -I_z$. The order of pulse elements combined with phase adjustments can also create pulses with $X$ and $Z$ rotation symmetries (table \ref{tab:sym}). Pulses can be created where the back half is computed from order/phase reversing the front half in Seedless by specifying \verb|SYMPR| in the header. 

Finally we  consider a pulse with time reflection symmetry where
\begin{equation}
V=V_1^{xyz} V_2^{x,y,z}...V_2^{x,y,z}V_1^{x,y,z}.
\end{equation}
The propagator of the same pulse but applied at the frequency $-\Omega$ will be 
\begin{equation}
\overline{V}=V_1^{x,y,-z} V_2^{x,y,-z}...V_2^{x,y,-z}V_1^{x,y,-z}.
\end{equation}
By comparing the results to the Hermitian conjugate of $V$, the two propagators are related by
\begin{equation}
Z(\pi) V^\dagger Z^\dagger(\pi) = \overline{V} .
\label{eqSym}
\end{equation}
The two propagators at $\pm \Omega$ will perform the following transforms
\begin{equation}
    \begin{split}
    \rho_s &\xrightarrow{V} \rho_f \\
    Z(\pi) \rho_f Z^\dagger(\pi) &\xrightarrow{V^\prime} Z(\pi) \rho_s Z^\dagger(\pi). \\
    \end{split}
\end{equation}
If the action of the pulse is $I_z\rightarrow -I_y$ at $+\Omega$ then at $-\Omega$ it will perform the transform $I_y\rightarrow I_z$. These are different state-to-state transforms and so time reflection symmetry isn't expected for frequency symmetric state-to-state excite/de-excite transformations. 

If instead the forward propagator performs a unitary $90^\circ$ rotation, then the action of both propagators at $\pm \Omega$ will be identical, and so time symmetry can be used to create a pulse that perform a frequency symmetric $90^\circ$ unitary rotation. This type of symmetry can be imposed in Seedless by specifying \verb|SYM| in the header. In practical GRAPE computations the best phase-only 90$_x^\circ$ pulses tend not to have this symmetry, as is also the case for commonly used pulses such as EBURP1~\cite{burp1991} and Q5~\cite{emsley1992}. Though the BEST-PC9 pulse has this symmetry by design~\cite{kupce1994,schanda2006}.

By contrast, if the state-to-state pulse is performing an inversion of either the $X$ or $Y$ axes then a pulse with time symmetry will perform the same transform at $\pm\Omega$ (e.g. REBURP~\cite{burp1991}). We find empirically that the `best' GRAPE frequency symmetric inversion pulses with a short duration naturally have this symmetry (Figure \ref{fig:180}), resembling BIP pulses~\cite{smith2001}. Computation can be made more efficient by enforcing this symmetry with \verb|SYM|. 

To understand the action of pulses performing unitary transforms, we analyse their propagator fidelities. The fidelities between a pulse propagator and a target $U$ at $\pm\Omega$ are
\begin{equation}
\begin{split}
{\mathcal{F}(\Omega)}&=\half{\mathrm{Tr}}\left[V(\Omega)\,U^\dag(\Omega)\right]\\
{\mathcal{F}(-\Omega)}&=\half{\mathrm{Tr}}\left[V(-\Omega)\,U^\dag(-\Omega)\right].\\
\end{split}
\end{equation}
For a symmetric frequency response we require $\mathcal{F}(-\Omega)=\mathcal{F}(\Omega)$. We can insert Equation \ref{eqSym}, note that the fidelities are real allowing us to take the adjoint, then cyclically permute, 
\begin{equation}
\begin{split}
{\mathcal{F}(-\Omega)}
&=\half{\mathrm{Tr}}\left[Z(\pi)\,V^\dag(\Omega)\,Z^\dag(\pi)\,U^\dag(-\Omega)\right]\\
&=\half{\mathrm{Tr}}\left[ U(-\Omega) Z(\pi)\,V(\Omega)Z^\dag(\pi)\right]\\
&=\half{\mathrm{Tr}}\left[V(\Omega)\,Z(\pi)U(-\Omega)Z^\dag(\pi)\right].
\end{split}
\end{equation}
Then $\mathcal{F}(-\Omega)=\mathcal{F}(\Omega)$ for a time symmetric pulse if the target operators at $\pm\Omega$ are related by
\begin{equation}
Z(\pi)\,U(-\Omega)\,Z^\dag(\pi)=U^\dag(\Omega).
\end{equation}
This is satisfied for schematic pulses where the central rotation is in the $xy$ plane, and if $a=b$ (which includes the case where $a=b=0$). The surrounding $z$-rotations will negate the $x$ and $y$ components of the central unitary, which if in the $xy$ plane has no $z$-component, so the overall effect is to invert the rotation as required. 

\section{Joining state-to-state rotations to obtain a unitary}
\label{app:Luy}
Luy et al. showed that two state-to-state rotations can be combined to create a unitary pulse~\cite{luy2005}. What are the formal requirements for this to work? Their proof starts from a target unitary propagator about the $x$ axis by an angle $\alpha$, that is split into two rotations.
\begin{equation}
\begin{split}
X(\alpha)&= X(\frac{\alpha}{2})X(\frac{\alpha}{2})\\
 &= X(\frac{\alpha}{2}) Y(\pi)  X(-\frac{\alpha}{2}) Y^\dagger(\pi)\\
 &= \textcolor{red}{X(\frac{\alpha}{2}) Y(\pi)  X^\dagger(\frac{\alpha}{2})} Y^\dagger(\pi)\\
 \end{split}
 \end{equation}
where in the second line we used $X(\frac{\alpha}{2})=Y(\pi)  X(-\frac{\alpha}{2}) Y^\dagger(\pi)$. The part in red describes a rotation of the $+Y$ axis about the $X$ axis by an angle of $\frac{\alpha}{2}$. We can evaluate this and identify a new rotation
\begin{equation}
 X(\alpha)= \textcolor{red}{U^{0,\cos\frac{\alpha}{2},\sin\frac{\alpha}{2}}(\pi)} Y^\dagger(\pi)
\end{equation}
Luy noted that it isn't essential to rotate $Y$ using the propagator $X(\frac{\alpha}{2})$. We can instead use any state-to-state propagator $V(\nu)$ that takes us from $I_y \rightarrow  I_y\cos\frac{\alpha}{2} + I_z\sin\frac{\alpha}{2}$,
\begin{equation}
\begin{split}
X(\alpha) &= \textcolor{red}{V(\nu) Y(\pi)  V^\dagger(\nu)} Y^\dagger(\pi).\\
 \end{split}
 \end{equation}
Finally, noting that we can order/phase reverse the elements of the first pulse to make a new one with the following symmetry $ Y(\pi)  V^\dagger(\nu) Y^\dagger(\pi) = V^\prime(\nu)$ (Appendix \ref{app:timePhaseRev},table \ref{tab:sym}), the overall effect will be a unitary rotation
\begin{equation}
\begin{split}
X(\alpha) &= V(\nu) V^\prime(\nu).\\
 \end{split}
 \end{equation}

The schematic propagator for this will be 
\begin{equation}
\begin{split}
X(\alpha) &= Z(a\Omega T) U^{x,y,z} U^{x,-y,z} Z(a\Omega T).\\
 \end{split}
 \end{equation}

Because the underlying state-to-state pulses $V(\nu)$ are not evolution controlled, $a$ will vary with $\Omega$ and so care must be taken when using these in a multi-spin context. In the special case of $X(\pi)$ unitary pulses, these can be constructed using either an excitation or a de-excitation state-to-state pulse, 
\begin{equation}
\begin{split}
X(\pi) &= V_{Y\rightarrow Z} V^\prime_{Y\rightarrow Z} \\
    &= V^\prime_{Z\rightarrow -Y} V_{Z\rightarrow -Y}. \\
    \end{split}
\end{equation}

What do we need in general to create a unitary rotation from state-to-state pulses? First, take a state-to-state transform $V$ that causes $\rho_s\rightarrow \rho_f$. A second pulse can be created by inverting the order of the first, and applying a $\pi$ rotation about the axis parallel with $\rho_s$ (e.g. the $X$, $Y$ or $Z$ axes, table \ref{tab:sym}). We construct the final pulse by applying the original immediately after the newly created symmetry partner. This will be a unitary rotation if rotating the back half of the pulse by $\pi$ about the symmetry axis generates the Hermitian conjugate of the front half. To take an example, using order/phase reversal fixes the symmetry axis to $Y$. A unitary pulse whose axis is in the XZ plane ($U^{x,0,z}(\psi)$) can then be created from a state-to-state pulse that starts on $I_y$ and has a final state 
\begin{equation}
    \rho_f = U^{x,0,z}(\frac{\psi}{2}) I_y U^{-x,0,-z}(\frac{\psi}{2})
\end{equation}

\section{Experimental methods}
\label{app:expmeth}

\begin{figure}
    \centering
    \includegraphics[trim={2.5cm 8.5cm 0 3cm},width=1.0\linewidth]{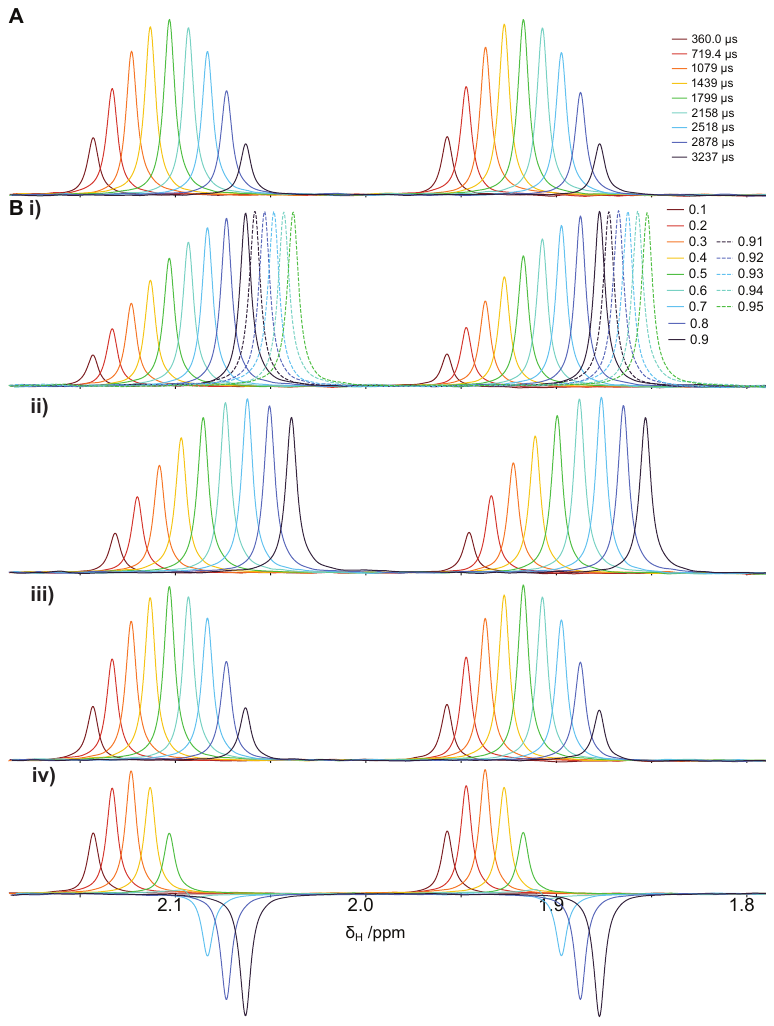} 
    \caption{Spectra of $^{13}$CH$_3$ L-methionine using a modified HSQC experiment (Figure \ref{fig:JINEPT}) where the effective scalar coupling time $\Delta$ was systematically varied. Spectra have been offset in a manner that is proportional to $\Delta$ for clarity. \textbf{A} Spectra acquired using rectangular pulses, where the evolution delay $\tau$ was varied. \textbf{B} Spectra acquired using the JINEPT element with evolution fraction $a$ varied as indicated and i) $T=946\;\mu$s , ii) $T=1285\,\mu$s, iii) $T=1799\,\mu$s, and iv) $T=2998\,\mu$s.}
    \label{fig:processed_spectra}
\end{figure}

To validate the JINEPT experiment, an HSQC experiment was adapted such that the first INEPT transfer could be performed either using rectangular pulses (HARD) or using a JINEPT (Figure \ref{fig:JINEPT}A). 1D HSQC spectra were acquired from a sample of $^{13}$CH$_3$ labelled L-methionine ($0.5$\,mM, in $99\;\%$ D$_2$O, $10$\,mM NaP$_i$, pH $7.4$). The period where scalar coupling was allowed to evolve, $\Delta$, was systematically altered in both cases. For HARD this was achieved by varying the delay $\tau$, and for the JINEPT, it was achieved by altering both the total duration of the \verb|a90x| pulses, $T$ and the period for which evolution is active, $a$. All spectra are shown (Figure \ref{fig:processed_spectra}). In all cases, signal intensity was well described by $A\sin (\Delta J \pi)$ (Figure \ref{fig:JINEPT} C), with a maximum at  $\Delta=\frac{1}{2J}$, leading to a fitted value of $J_{CH}$ of 138.6\,Hz (Figure \ref{fig:JINEPT}C). The agreement between the two methods demonstrates that scalar coupling can be precisely controlled using evolution controlled pulses.

Spectra were acquired on a Bruker Avance III HD 750\,MHz instrument at 25$^\circ$C. Each spectrum comprised 16 transients each with 19,530 complex points with a sweep width of $9765.625$\,Hz, each with an acquisition time of $1$ second. The relaxation delay was set to $1.5$ seconds. The carrier frequencies used for $^1$H and $^{13}$C were 5\,ppm and 13\,ppm respectively, which are close to the H/C resonance frequencies of methionine such that the performance of the rectangular pulses was maximised. The 90$^\circ$ rectangular pulses were calibrated to require $7.84\,\mu s$ and $21.7\,\mu s$ for $^1$H and $^{13}$C respectively at maximum amplifier power.

FIDs were processed with NMRPipe~\cite{NMRPipe_Delaglio1995}. They were zero filled to double the number of points, a 5\,Hz exponential window function was applied, the first point was scaled by 0.5 and a Fourier transform was applied. A manual zeroth order phase correction was applied before extraction of the region 0 to 5\,ppm for analysis. The $^1$H resonance was split into two by the $^{13}$C coupling, leading to two resonances ($2.14$ and $1.96$\,ppm). The maximum intensity of both were determined, allowing a mean and standard deviation to be evaluated for each value of $\Delta$ (Figure \ref{fig:JINEPT}C).

For HARD, 9 values of $\tau$ was measured, varied in the range $359.7\,\mu$s to $3237\,\mu$s. For the JINEPT, \verb|a90x| pulses (and their time- and phase-reversed counterparts) were created with four different durations, $T=$ 2998, 1799, 1285 and 946\,$\mu$s. The values of $a$ for which maximum transfer was achieved at each duration were 0.3, 0.5, 0.7 and 0.9\,5 for the 4 durations, where $aT= 899.5$ $\mu$s, and $\Delta=\frac{1}{2J}=4aT=3597\;\mu s$. 9 difference values of $a$ were tested for each duration, selected from the range 0.1 to 0.9. In addition, for the pulse of 946 \,$\mu$s, values of 0.90 to 0.95 in increments of 0.01 were analyzed, to question performance close to the maximum, and where pulse infidelity starts to increase significantly, reflecting the optimization challenge of allowing evolution for $>$90\% of the duration of the pulse. A total of 41 a90x pulses were generated to test 41 JINEPT elements (Figure \ref{fig:JINEPT}C).

The Seedless \verb|a90x| pulses for the JINEPT and their order/phase reversed counterpart (automatically generated by specifying \verb|WritePR| in the input script) were calculated for a B$_1$ field of $15,723$\,Hz (equivalent for a 90$^\circ$ pulse of length $15.9\,\mu$s). This corresponds to half of the $B_1$ of the maximum, which reduces the power requirements by a factor of 4. As the JINEPT experiment is continuously exciting for several milliseconds, this kept the overall power requirements well within the hardware specifications. In the Seedless calculation, the maximum number of iterations was set to 30,000, and $B_1$ robustness was achieved by including fields at 0.93, 1.0 and 1.05 of the central value weighted $\frac{1}{4}$,$\frac{1}{2}$,$\frac{1}{4}$ respectively~\cite{seedless}. 100 spins were included, with evolution frequencies spaced linearly in the range 0 to 5\,ppm, with the pulse carrier set to\,5 ppm. A total of 41 unique \verb|a90x| pulses were produced with varying $T$ and $a$, as described above.

\section{Seedless scripts}
\label{app:SeedScript}
Seedless can be downloaded from \verb|https://seedless.chem.ox.ac.uk|. A complete manual is provided together with demonstration input files allowing examples from this work to be easily replicated and used as a template for new  applications. All GRAPE and non-GRAPE pulses in this work were computed using Seedless (Table \ref{tab1}, Figures \ref{fig:90},\ref{fig:180}). For the GRAPE pulses, the following restraints for the target bands were used (where just one band of chemical shifts was specified)
\begin{verbatim}
Targets: 
a90xb
0.5O;0.25B;0.5O 
90x
Iz -Iy
Iz -0.2OIy
180x
Iz -Iz
Iz Iex
\end{verbatim}

These compute (respectively) \verb|a90xb| (used in conjuction with \verb|evAlpha| and \verb|evBeta| to set $a$ and $b$, specified in the header), \verb|a90xb| (an explicit description, setting $a=b=0.5$), \verb|90x|, two evolution controlled state to state ($b=0$, $b=0.2$), and three pulses for which $a$ and $b$ are uncontrolled, a \verb|180x|, an inversion and an \verb|XYcite|. 

To compute an evolution controlled refocusing pulse, \verb|a180xb|, specify \verb|REBURP| in the header, set \verb|evAlpha=a|, \verb|evBeta=0| (pre- and mid- evolution delays), and set \verb|180x| as the target (see example script below).

The non-GRAPE pulses (Table \ref{tab1}) were computed using the following commands. A total duration and number of points is specified, and the $B_1$ field is then optimised (\verb|Opt|) to find the value that performs the desired transformation on resonance. Note one `make' command should be specified at a time otherwise the pulses will concatenate. Further, \verb|NoOpt| should be set so that the optimiser does modify the pulse after setting it up, and \verb|CalcInit| will perform the Seedless cost function calculation for the final report. For excitation/de-excitation,
\begin{verbatim}
Make HARD 2 1000 Opt Iz Iy -1
Make EBURP1 1000 2000  Opt Iz Iy -1
Make PC9 500 1000 Opt Iz Iy -1
Make BESTPC9 2000 1000 Opt Iz Iy -1
Make Q5 1000 2000 Opt Iz Iy -1
\end{verbatim}

and for 180$^\circ$ rotation,
\begin{verbatim}
Make HARD 2 2000 Opt Iz Iz -1
Make REBURP1 1000 2000  Opt Iz Iz -1
Make Q3 1000 2000 Opt Iz Iz -1
Make CHIRP 1000 2000 sweep 20000 n 5 SetB1 4060 tag 2 Opt Iz Iz -1 wHmax 10000
\end{verbatim}

The syntax requires an identifier (which can also be a file location), a number of points and a duration (in $\mu$s). The $B_1$ field for optimal performance is then numerically determined according to the restraints specified as a starting operator (\verb|Ix,Iy,Iz|), a finishing operator (\verb|Ix,Iy,Iz|) and value required for the finishing operator. The $B_1$ field values obtained in pulses described in  this work (table \ref{tab1}) were 0.25 KHz (HARD90), 1.861 KHz (EBURP1) 2 KHz (PC9), 1 KHz (BESTPC9), 2.27 KHz (Q5),  0.25 KHz (HARD180), 3.12 KHz (REBURP), 1.65 KHz (Q3), 3.76 KHz (CHIRP).

To compute the amplitude (0--1) and phase (degrees) of CHIRP pulses~\cite{chirp1989}, for a sweep width ($s_w$), total duration ($T$), number of points ($N$), amplitude scaling factor ($n$) and $g_i=\frac{iT}{N}$, the first $\frac{N}{2}$ points can be computed from 
\begin{equation}
\begin{split}
   A_i &= \sin\left(\frac{\pi g_i }{2 n T}\right) \\
\phi_i &= \left(1 - \frac{g_i}{ T} \right) g_i\frac{s_w}{2}\, 360 \\
\end{split}
\end{equation}
and the back half is a time reflection of the front. This will perform an adibatic sweep from $-\frac{s_w}{2}$ to $\frac{s_w}{2}$. The sweep width of 20\,KHz was chosen to ensure adiabacicity and the field of 3.76\,KHz was optimised by Seedless to be a reasonable minimum value to reliably achieve a $B_1$ robust inversion. This has been implemented in Seedless via the syntax above. The schematic form shows the pre- and post- delays varying with offset in a linear manner that matches the sweep, where the pre- delay at the start of the sweep being zero, and the post- delay being close to 1, and vice-versa by the end of the pulse. The CHIRP pulse is not band-schematic, and will result in offset dependent evolution times when used in e.g. INEPT sequences.

\begin{figure}
\includegraphics[trim={0cm 0cm 0 0},width=0.8\linewidth]{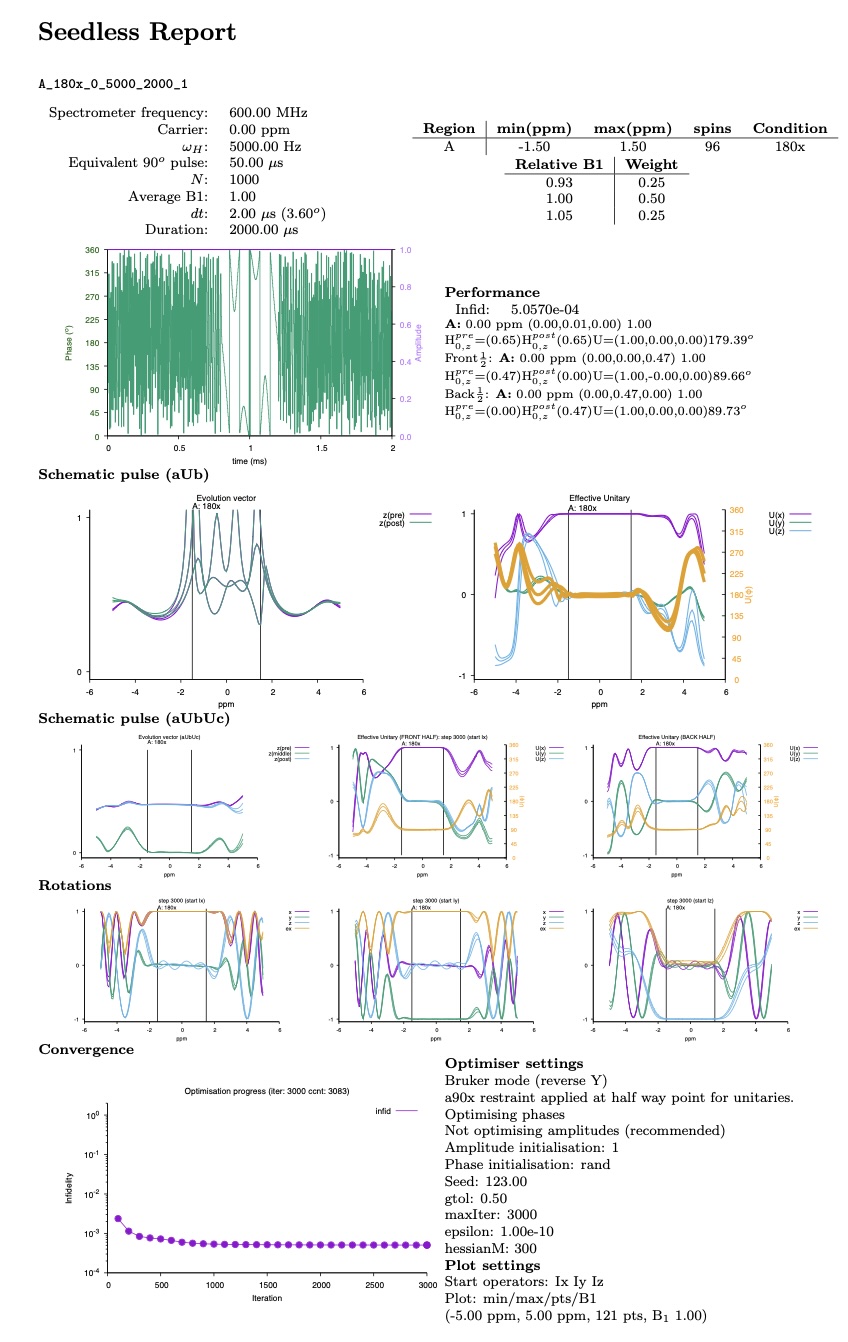}
\caption{A summary report generated automatically by Seedless showing the computation of an a180xa pulse. Schematic pulse analysis is turned on by specifying EVOLVE. Because this is a 180$^\circ$ pulse, the evolution parameters are unreliable as discussed in the text. Specifying HALF analyses both halves of the pulse, allowing determination of the pre, mid and post evolution delays. The pulse was not constrained to have order/phase reversal symmetry, this symmetry in the final pulse emerged from optimisation.}
\label{fig:sumrep}
\end{figure}

The following is a complete sample input script for an \verb|a180xa| evolution controlled refocusing pulse. The associated report is shown in Figure \ref{fig:sumrep}.
\begin{small}
\begin{verbatim}
ncpus 16   # cpu count
frq 600    # Larmor frequency (MHz)
IterShow 100   #show progress every this many steps.
maxIter 3000   #maximum number of iterations 
MakeSummary    # Make summary pdf
REBURP     #turn on REBURP mode
EVOLVE     #turn on schematic pulse analysis
HALF       #turn on schematic pulse analysis at halfway 
evAlpha 0.475 #set evolution parameter a
evBeta 0.0    #set evolution parameter b
Bruker        #write out pulses in Bruker format
RF: (B1/weight)  #B1 robustness
 0.93	0.25
 1.0	0.5
 1.05	0.25
Plot: (Min/Max/Num)  #spins to plot
 P -5 5   121
SpinSystem: (Min/Max/Num/Target)  #spins to optimise
 A  -1.5 1.5  96 
Carriers: #center of spectrum (ppm)
 0.0
wmH:      #B1 field (Hz)
 5000   
Durations: #Total duration(s) number of points
 2000E-6 1000
Targets:   #retraints for the bands
 180x
\end{verbatim}
\end{small}

\noindent

\end{document}